\documentclass[a4paper,fleqn,usenatbib]{mnras}
\usepackage{epsf,palatino,url,graphicx,wrapfig,array,setspace,amsmath,amssymb,fancyhdr,multirow,lscape,appendix,rotating,txfonts}
\usepackage{graphics,epsfig,txfonts}
\bibpunct{(}{)}{;}{a}{}{,}

\usepackage{longtable}
\usepackage{amssymb}
\usepackage{xcolor}
\usepackage{color}
\usepackage{lipsum}
\usepackage{textcomp}
\usepackage{threeparttable}
\usepackage{enumerate}

\usepackage{placeins}
\usepackage{float}

\usepackage[normalem]{ulem}
\usepackage{soul}
\usepackage{blindtext}
\usepackage[hyphenbreaks]{breakurl}

\newcommand{\oergcm}[1]{$10^{#1}$ erg cm$^{-2}$ s$^{-1}$}
\newcommand{\ergs}[1]{$\times 10^{#1}$ erg s$^{-1}$}
\newcommand{\oergs}[1]{$10^{#1}$ erg s$^{-1}$}

\newcommand{\ohcm}[1]{$10^{#1}$ cm$^{-2}$}

\newcommand{\nh}{N$_{\rm H}$}

\newcommand{\ltsima}{$\buildrel < \over \sim$}
\newcommand{\lsim}{\lower.5ex\hbox{\ltsima}}
\newcommand{\gtsima}{$\buildrel > \over \sim$}
\newcommand{\gsim}{\lower.5ex\hbox{\gtsima}}

\newcommand{\swift}{{\it Swift}\xspace}

\newcommand{\xmm}{{\it XMM-Newton}\xspace}

\newcommand{\canda}{XMMU\,J053108.3-690923\xspace} 
\newcommand{\candb}{XMMU\,J053320.8-684122\xspace} 

\newcommand{\eqb}{\begin{eqnarray}}
\newcommand{\eqe}{\end{eqnarray}}

\begin{document}

\title[Two new HMXB in the LMC]{Identification of two new HMXBs in the LMC: a $\sim$2013 s pulsar and a probable SFXT}
\author[G. Vasilopoulos et al.]{G.~Vasilopoulos,$^1$\thanks{E-mail: gevas@mpe.mpg.de} C.~Maitra,$^1$ F.~Haberl,$^1$ D.~Hatzidimitriou,$^{2,3}$ M.~Petropoulou$^4$\\
$^1$Max-Planck-Institut f\"ur extraterrestrische Physik,Giessenbachstra{\ss}e, 85748 Garching, Germany \\
$^2$ National and Kapodistrian University of Athens, Department of Physics, Section of Astrophysics, Astronomy and Mechanics, Panepistiniopolis,  GR 15783 Zografos, Greece \\
$^3$  IAASARS, National Observatory of Athens, GR 15236 Penteli, Greece \\
$^4$  Department of Astrophysical Sciences, Princeton University, 4 Ivy Lane, Princeton, NJ 08544, USA
}
\date{Accepted MNRAS}

\maketitle

\begin{abstract}
We report on the X-ray and optical properties of two high-mass X-ray binary systems located in the Large Magellanic Cloud (LMC).
Based on the obtained optical spectra, we classify the massive companion as a supergiant star in both systems. 
Timing analysis of the X-ray events collected by \xmm revealed the presence of coherent pulsations (spin period $\sim$2013~s) for \canda and fast flaring behaviour for \candb.
The X-ray spectra of both systems can be modelled sufficiently well by an absorbed power-law, yielding hard spectra and high intrinsic absorption from the environment of the systems.
Due to their combined X-ray and optical properties we classify both systems as SgXRBs: the 19$^{\rm th}$ confirmed X-ray pulsar 
and a probable supergiant fast X-ray transient in the LMC, the second such candidate outside our Galaxy.
\end{abstract}

\begin{keywords}
galaxies: individual: Large Magellanic Cloud --
         X-rays: binaries --
         stars: neutron --
         pulsars: individual:
\end{keywords}

\section{Introduction}
\label{sec-intro}

X-ray binaries (XRBs) are close binary systems composed of a compact object and an optical companion star that is usually undergoing nuclear burning.
They can be classified into different categories based on the type of the compact star (i.e. black hole, neutron star, white dwarf) or the 
optical companion (low-mass or high-mass star). For black-hole (BH) or neutron-star (NS) binaries it has been shown that the number of low-mass X-ray binaries within a galaxy scales with the galaxy's mass \citep{2004MNRAS.349..146G} while the number of high-mass X-ray binaries (HMXBs) is proportional to the recent star formation rate of the galaxy.
HMXBs can be divided into two categories depending on the luminosity class of the donor star \citep[for a detailed review see ][]{2011Ap&SS.332....1R}: 
Supergiant X-ray binaries (SgXRBs) for luminosity class I-II stars and Be/X-ray Binaries (BeXRBs) when the optical companion is a dwarf, sub-giant or giant OBe star. 

It has been established that HMXB systems show strong aperiodic variability \citep{1990A&A...230..103B}.
On the one hand, BeXRB systems are transient systems exhibiting periodic Type-I outbursts near the NS periastron passage and sporadically major Type-II outbursts associated with periods of increased accretion probably due to the formation of warped Be-disks \citep{2013PASJ...65...41O}. 
Wind-fed systems like SgXRBs, on the other hand, have a fairly constant luminosity with typical hourly variability as a result of clumped stellar wind accretion \citep[e.g.][]{2012MNRAS.421.2820O,2014A&A...563A..70M}.

The standard classification is challenged by newly discovered systems that do not seem to fit into this scheme. 
Within the last decade a new population of HMXBs \citep{2016ApJS..223...15B}, which is demographically different to the previously studied HMXB population, has been discovered by the International Gamma-ray Astrophysics Laboratory (INTEGRAL).
This new population of HMXB systems is usually characterized by large obscuration, while it exhibits extreme flaring activity on timescales of a few hours. 
Both properties led to their identification as supergiant fast X-ray transients (SFXT) \citep{2006ESASP.604..165N,2017arXiv171003943S}. It has been proposed that the SFXT phenomenon is fairly common \citep{2014A&A...568A..76D}, but the detection of the entire population in the Galaxy is hindered by their transient properties and the large Galactic absorption. Given the above observational biases, it is no surprise that only one extragalactic SFXT has been detected so far \citep[IC\,10\,X-2, ][]{2014ApJ...789...64L}.

\begin{table*}
\caption{X-ray observations log}
\begin{center}
  \scalebox{0.9}{
 \begin{threeparttable}
\begin{tabular}{lccccrrr}
\hline\noalign{\smallskip}
     \multicolumn{1}{c}{Observation} &
     \multicolumn{1}{c}{ObsID} &
     \multicolumn{1}{c}{Date} &
     \multicolumn{1}{c}{Instrument} &
     \multicolumn{1}{c}{Mode $^a$} &
     \multicolumn{1}{c}{Net Exp} &  
     \multicolumn{1}{c}{rate} &
     \multicolumn{1}{c}{L$_x$ $^b$} \\
     
     \multicolumn{1}{c}{Target/Observatory} &
     \multicolumn{1}{l}{} &
     \multicolumn{1}{c}{MJD} &
     \multicolumn{1}{l}{} &
     \multicolumn{1}{l}{} &
     \multicolumn{1}{c}{[ks]} &   
     \multicolumn{1}{c}{counts/s} &
     \multicolumn{1}{c}{\oergs{35}} \\

     \noalign{\smallskip}\hline\noalign{\smallskip}

\hline\noalign{\smallskip}
\canda \\
\xmm     & 0690744601  &  2012-10-10    &  EPIC-pn    &  ff--thin   &  40  &  0.20$\pm$0.03 & 6.4\\
         &             &                &  EPIC-MOS1  &  ff--medium &  41  &  0.071$\pm$0.002 & \\
         &             &                &  EPIC-MOS2  &  ff--medium &  41  & 0.073$\pm$0.002  & \\
\swift   & 00032616001 &  56237.55421   &  XRT        &  pc         &   4  &  $<$0.0025 & $<$0.8 \\
         & 00045434001 &  56241.56838   &  XRT        &  pc         & 0.2  &  $<$0.027 & $<$8.8\\
         & 00045434003 &  56247.04019   &  XRT        &  pc         & 1.3  &  $<$0.015 & $<$4.8\\
         & 00045436002 &  56247.17643   &  XRT        &  pc         & 1.4  &  $<$0.007 & $<$2.3\\
         & 00045436004 &  56257.85074   &  XRT        &  pc         & 0.8  &  $<$0.008 & $<$2.6\\
         & 00045463001 &  56268.25574   &  XRT        &  pc         & 2.5  &  $<$0.006 & $<$1.95\\
         & 00032616002 &  56967.49831   &  XRT        &  pc         & 0.8  &  0.015$\pm$0.005 & 4.9\\
         & 00032616003 &  56969.69179   &  XRT        &  pc         & 0.8  &  $<$0.017 & 5.2\\

\hline\noalign{\smallskip}
\candb \\
\xmm     & 0690743801  &  2012-04-27    &  EPIC-pn    &  ff--thin   &  26  & 0.137$\pm$0.004  & 2.8\\
         &             &                &  EPIC-MOS1  &  ff--medium &  28  & --  & \\
         &             &                &  EPIC-MOS2  &  ff--medium &  28  &  0.040$\pm$0.002 & \\         
\xmm     & 0690743701  &  2012-08-23    &  EPIC-pn    &  ff--thin   &  30  &  -- & 0.21\\
         &             &                &  EPIC-MOS1  &  ff--medium &  31  &   0.004$\pm$0.001 & \\
         &             &                &  EPIC-MOS2  &  ff--medium &  31  &  0.003$\pm$0.001 & \\
\swift   & 00045431001 & 55999.30753    &  XRT        &  pc         &  2.0 & 0.013$\pm$0.003 & 3.5\\
         & 00032310001 & 56006.38836    &  XRT        &  pc         &  2.9 & 0.021$\pm$0.003 & 5.7\\
         & 00045431003 & 56067.34919    &  XRT        &  pc         &  0.3 & $<$0.04 & $<$11.\\
         & 00032310002 & 56967.76268    &  XRT        &  pc         &  1.0 & $<$0.006 & $<$1.6\\
         & 00045567001 & 56376.06192    &  XRT        &  pc         &  2.6 & $<$0.009 & $<$2.5\\
         & 00045566002 & 56379.25832    &  XRT        &  pc         &  1.2 & $<$0.009 & $<$2.5\\
         & 00032310003 & 57109.74894    &  XRT        &  pc         &  0.2 & $<$0.15 & $<$40.\\

\hline\noalign{\smallskip}
\end{tabular}
\tnote{a} Instrument setup modes. For \xmm a full-frame (ff) mode with thin or medium filter was used. For \swift/XRT the photon counting (pc) operation mode was used.   \\
\tnote{b} To convert \swift/XRT count rates to luminosities we used the spectral parameters obtained from the fit to the \xmm data. The conversion factor is 3.25$\times 10^{37}$ erg s$^{-1}$ (c/s)$^{-1}$ for \canda (see case I in Table \ref{tab:spectra}) and 2.73$\times 10^{37}$ erg s$^{-1}$ (c/s)$^{-1}$ for \candb.\\
\end{threeparttable}
 }
\end{center}
\label{tab:log}
\end{table*}

In the case of the Magellanic Clouds (MCs)  all but one of the known XRB systems \citep[i.e. LMC\,X-2 is low-mass X-ray binary][]{2003ApJ...590.1035S} are HMXBs in nature \citep[e.g.][]{2010ApJ...716L.140A,2016MNRAS.459..528A}. The Small Magellanic Cloud (SMC) alone harbours about 120 HMXBs of which 64 are confirmed BeXRB pulsars and only one is a confirmed SgXRB \citep{2016A&A...586A..81H,2017MNRAS.470.4354V,2014MNRAS.438.2005M}. The population of BeXRBs was shown to be associate with the recent star formation history of the galaxy that peaks at $\sim$40\,Myr ago; this is also the time needed for the evolution of a high mass binary system into a BeXRB \citep{2010ApJ...716L.140A}.  
The Large Magellanic Cloud (LMC) shows evidence for more recent star formation than the SMC \citep{2009AJ....138.1243H}, and  could be thus a good candidate for detecting more SgXRB systems. 
Our knowledge of the X-ray population of the LMC is much less complete than that of the SMC. 
This is mainly due to its larger projected angular size in the sky that hinders extensive monitoring observations, while deep X-ray observations with modern observatories can only cover a small fraction of the galaxy.
Nevertheless, the LMC is known to harbour four SgXRB systems out of the $\sim$40 HMXB candidates \citep{2016MNRAS.459..528A}.
The recent \xmm LMC survey (PI: F.~Haberl) aimed to address the lack of deep X-ray observations in the LMC, by homogeneously covering a $\sim$15 square degree area of the nearby galaxy down to a limiting point source luminosity of 2\ergs{33}.

In this work, we report on the properties of two new HMXBs in the LMC. 
Although one of the objects has already been reported as a HMXB candidate \citep[][]{2012ATel.3993....1S}, here we expand upon the findings of the initial announcement by presenting a thorough analysis and review of the system's properties.
Both systems were detected at a moderate luminosity of $\sim$\oergs{35} during the \xmm LMC survey allowing a detailed spectral and temporal study of their properties.  
The study of their long-term X-ray variability is complemented by \swift/XRT observations.
Finally, the physical characteristics of the donor stars are derived with the aid of follow-up optical spectroscopic observations using the FEROS spectrograph on the ESO/MPG\,2.2\,m telescope.


\section{Observations}
\label{sec-observations}


\subsection{X-ray data}
The two targets were serendipitously discovered in X-rays by \swift and \xmm.
\candb was firstly identified as a HMXB candidate following a \swift/XRT detection \citep[Swift\,J053321.3-684121:][]{2012ATel.3993....1S} during a \swift/UVOT survey observation of the LMC (PI: S. Immler) performed on 2012 March 13. It was later serendipitously detected during two \xmm pointings, performed on April 27 and August 23 of 2012, as part of the \xmm large survey of the LMC.  \canda was identified as a HMXB candidate in the data of an \xmm observation performed on October 10 2012 as part of the LMC survey. Additional \swift/XRT observations were obtained at the position of each source to monitor their long-term X-ray variability. A detailed log of the \xmm and \swift observations that were analysed for the current work is presented in Table~\ref{tab:log}.

\xmm/EPIC \citep{2001A&A...365L..18S,2001A&A...365L..27T} data were processed using the latest \xmm data analysis software SAS, version 16.1.0\footnote{Science Analysis Software (SAS): \url{http://xmm.esac.esa.int/sas/}}. Observations were inspected for high background flaring activity by extracting the high-energy light curves (7.0\,keV$<$E$<$15\,keV for both MOS and pn detectors) with a bin size of 100\,s. Event extraction was performed with SAS task \texttt{evselect} using the standard filtering flags (\texttt{\#XMMEA\_EP \&\& PATTERN<=4} for EPIC-pn and \texttt{\#XMMEA\_EM \&\& PATTERN<=12} for EPIC-MOS).
Circular regions with a radius of 20\arcsec~and 50\arcsec~ were used for the source and background extraction to maximize the number of extracted photons.
For the spectral analysis the SAS tasks \texttt{rmfgen} and \texttt{arfgen} were used to create the redistribution matrix and ancillary file.
The grouped spectra were binned to achieve a minimum signal-to-noise ratio of five for each bin.
For the timing analysis the event arrival times were corrected to the solar-system barycentre by using the SAS task {\tt barycen}. 

The \swift/XRT data were analysed following standard procedures described in the \swift data analysis guide\footnote{\url{http://www.swift.ac.uk/analysis/xrt/}} \citep{2007A&A...469..379E}. We used the {\tt xrtpipeline} to generate the \swift/XRT products. Because of the low luminosity of the systems in most observations, we only performed a simple source detection and position determination using a sliding-cell detection algorithm implemented by {\tt XIMAGE}\footnote{\url{https://heasarc.gsfc.nasa.gov/xanadu/ximage/ximage.html}}. 
For non-detections we estimated the 3$\sigma$ upper limits using a Bayesian method introduced by \citet{1991ApJ...374..344K}.

\begin{figure}
      \includegraphics[width=0.95\columnwidth,clip=,bb= 77 170 480 570]{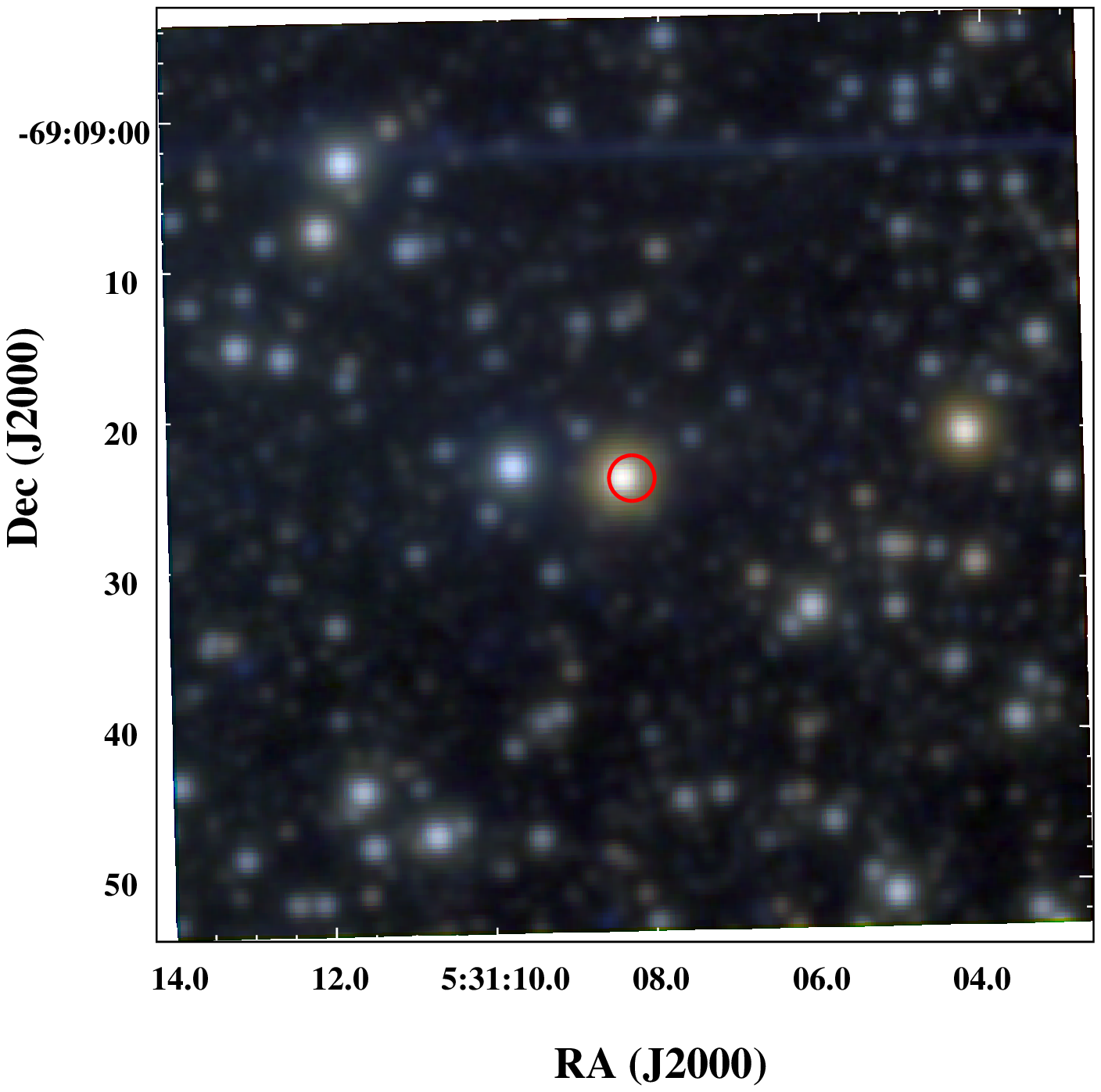}\\
      \includegraphics[width=0.95\columnwidth,clip=,bb= 77 170 480 570]{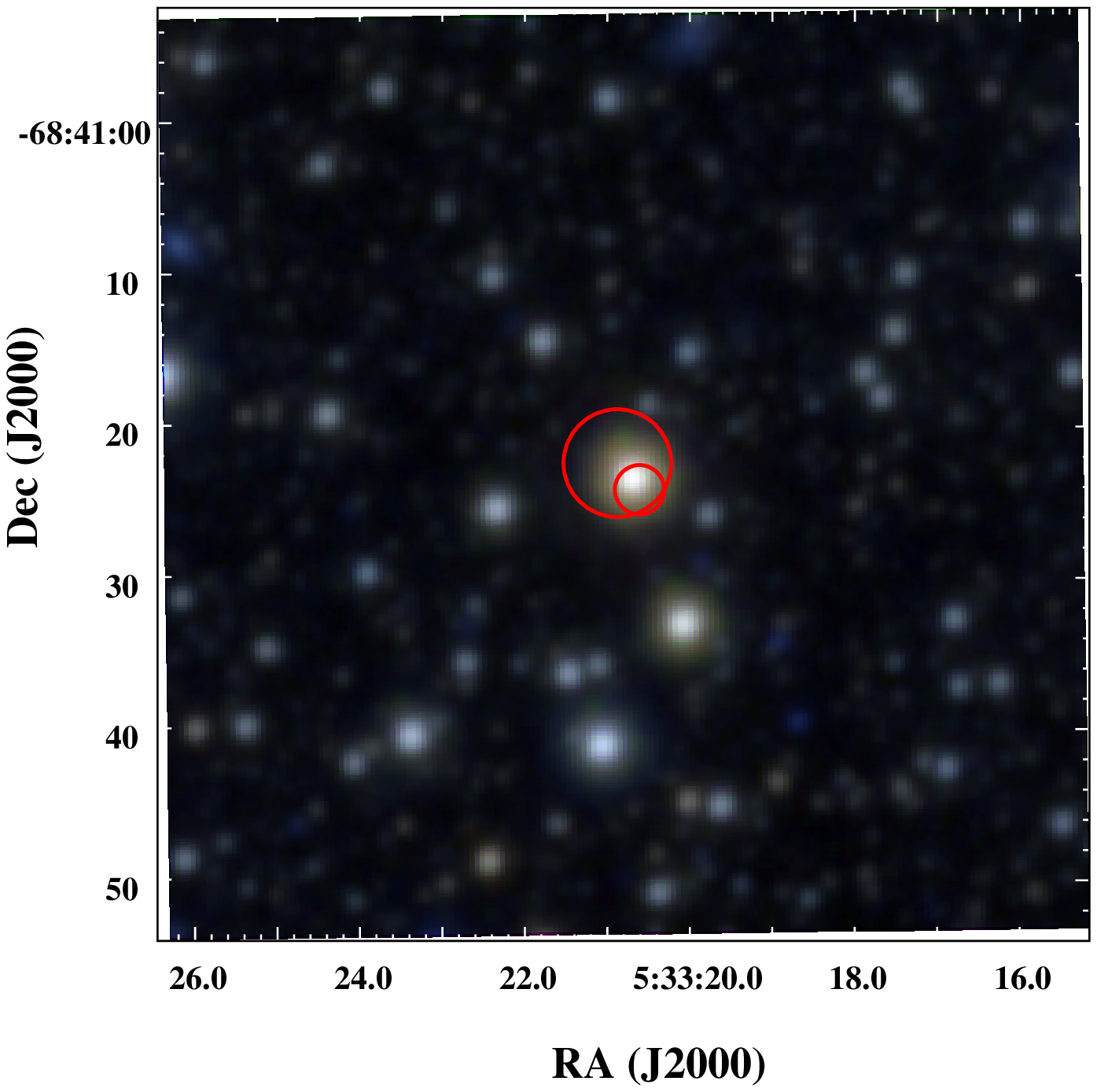}
\caption{ 
Vista images of the regions around \canda (top) and \candb (bottom). Each image covers a 1\arcmin\ x 1\arcmin\ sky area, with 
RGB colours based on the J, H, and K Vista filters respectively. Red circles are centred on the inferred \xmm positions with a radius corresponding to 3$\sigma$ confidence (including statistical and systematic errors). The two circles in the case of \candb correspond to two different \xmm observations. During the April 2012 observation the source was brighter and the position more accurate (smaller circle).}
  \label{fig:finding}
\end{figure}

\subsection{Optical data}

\begin{table}
 \centering 
 \caption {Optical counterparts}
 	\label{tab:opt}
\begin{threeparttable}
\begin{tabular}{lcc}
\hline\noalign{\smallskip}
X-ray  &  \canda & \candb  \\
\hline\noalign{\smallskip}
[M2002]$^a$  & LMC 150004 & LMC 157075  \\
\noalign{\smallskip}
V & 13.68$\pm$0.01 & 12.72$\pm$0.01 \\
\noalign{\smallskip}
B & 13.76$\pm$0.01& 12.63$\pm$0.01 \\
\noalign{\smallskip}
U & 12.89$\pm$0.01 & 11.67$\pm$0.01 \\
\noalign{\smallskip}
R & 13.6$\pm$0.01& 12.71$\pm$0.01\\
\hline\noalign{\smallskip}
[AAVSO]$^b$  & - & - \\
\hline\noalign{\smallskip}
g' &13.654$\pm$0.047 & 12.548$\pm$0.010\\
\noalign{\smallskip}
r' & 13.769$\pm$0.077& 12.792$\pm$0.040 \\
\noalign{\smallskip}
i' & 13.872$\pm$0.066& 12.968$\pm$0.083  \\
\hline\noalign{\smallskip}
\end{tabular}
\tnote{a} Catalogue number and photometric magnitudes from \citet{2002ApJS..141...81M}, typical photometric uncertainties are below 0.01 mag. \\
\tnote{b} AB magnitudes using Sloan filters \citep{2015AAS...22533616H}\\
\end{threeparttable}
\end{table}
Based on the accurate X-ray positions as determined by \xmm we identified the most probable counterpart in the LMC field.
A high-mass star is located within the X-ray error circle of each system (see Table \ref{tab:opt}). 
We consider these as the most likely counterparts, since it is very unlikely to find low-mass X-ray binary systems in the LMC \citep{2004MNRAS.349..146G} due to its recent star formation history and its total mass \citep{2009AJ....138.1243H,2016MNRAS.459..528A}.
The X-ray detected systems are not background Active Galactic Nuclei (AGN), given their large X-ray variability \citep[AGN have typical normalized excess variance $<$1,][]{2014ApJ...781..105L} and their hard X-ray spectra  (see \S \ref{results}). 
In Fig.~\ref{fig:finding} we present a finding chart for each system. 
The images were constructed using J, H and K publicly available VISTA images\footnote{http://horus.roe.ac.uk/vsa/index.html}.

Optical spectra were obtained using ESO's Fibre-fed Extended Range Optical Spectrograph \citep[FEROS,][]{1999Msngr..95....8K} mounted on the ESO/MPG\,2.2\,m telescope at La Silla (Chile).
Observations were performed using the MPE high-energy group internal observing time. Observations were scheduled around full moon periods when observing conditions were not optimal for photometric observations with GROND \citep{2008PASP..120..405G}, that is the telescope's main instrument.
The targets were observed during non-photometric nights during 2014 November 7, with $\sim$1\,h of total exposure for each target.
Reduction of the FEROS spectra was performed with the ESO Data Reduction System (DRS), provided to the FEROS users \footnote{\url{www.eso.org}}. 

\begin{figure*}
  \resizebox{\hsize}{!}{
     \includegraphics[angle=0,clip=]{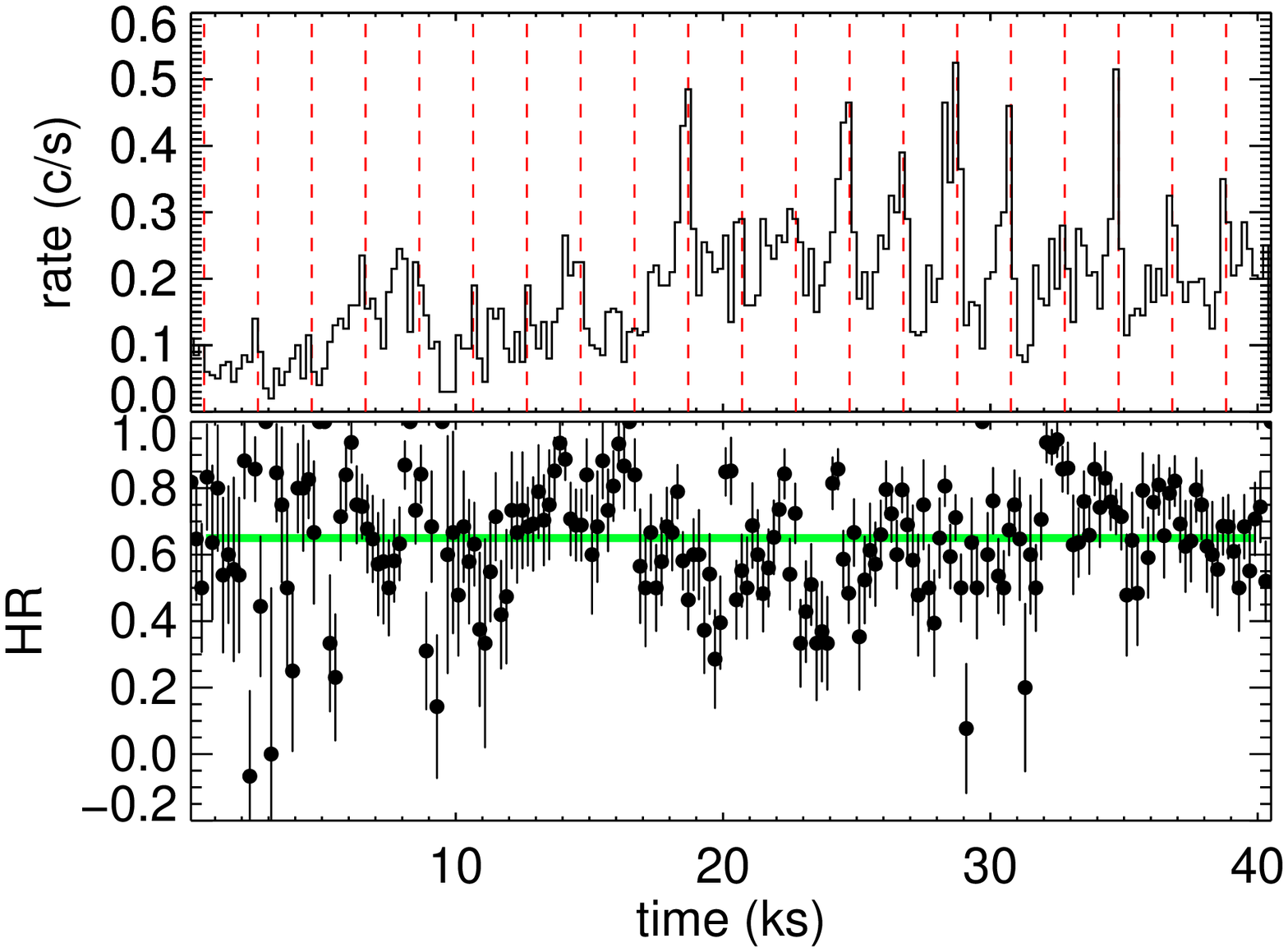}
     \includegraphics[angle=0,clip=]{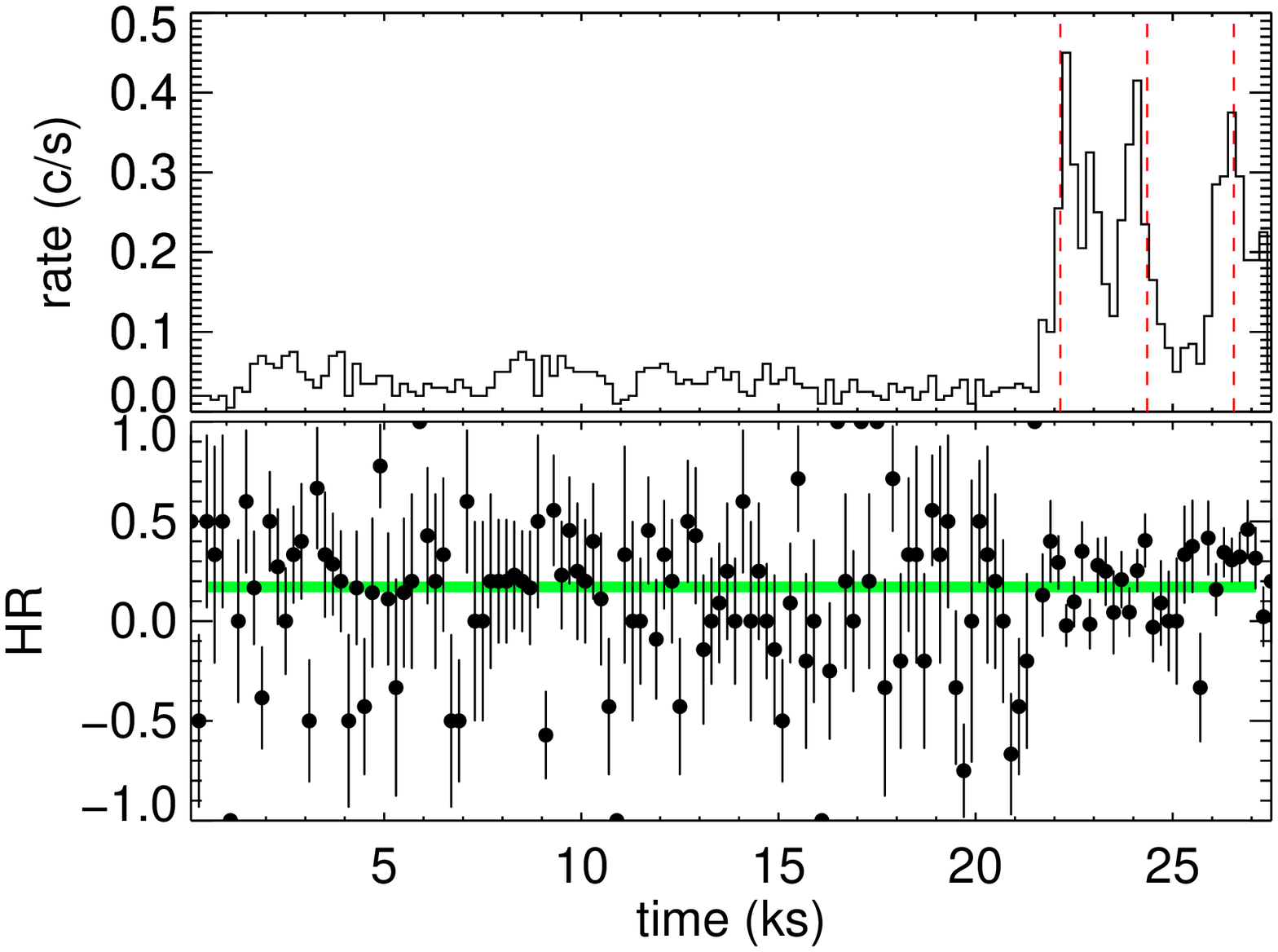}}
  \caption{\xmm/EPIC X-ray (0.3-10.0 keV) light curves with 200\,s binning, and the corresponding hardness ratios using the 0.3-2.0 keV and the 2.0-10.0 keV bands.
  \canda (left) shows clear evidence of periodic bursts every $\sim$2013\,s (red vertical lines).
  During the April 2012 observation (obsid: 0690743801) \candb (right) remained at a relatively constant luminosity ($\sim$0.035 counts s$^{-1}$) for the first $\sim$22\,ks of the exposure, while exhibiting three consecutive bursts separated by $\sim$2200\,s at the end of the exposure.   
  } 
  \label{fig:601_LC}
\end{figure*}

\begin{figure*}
  \resizebox{\hsize}{!}{
     \includegraphics[angle=0,clip=]{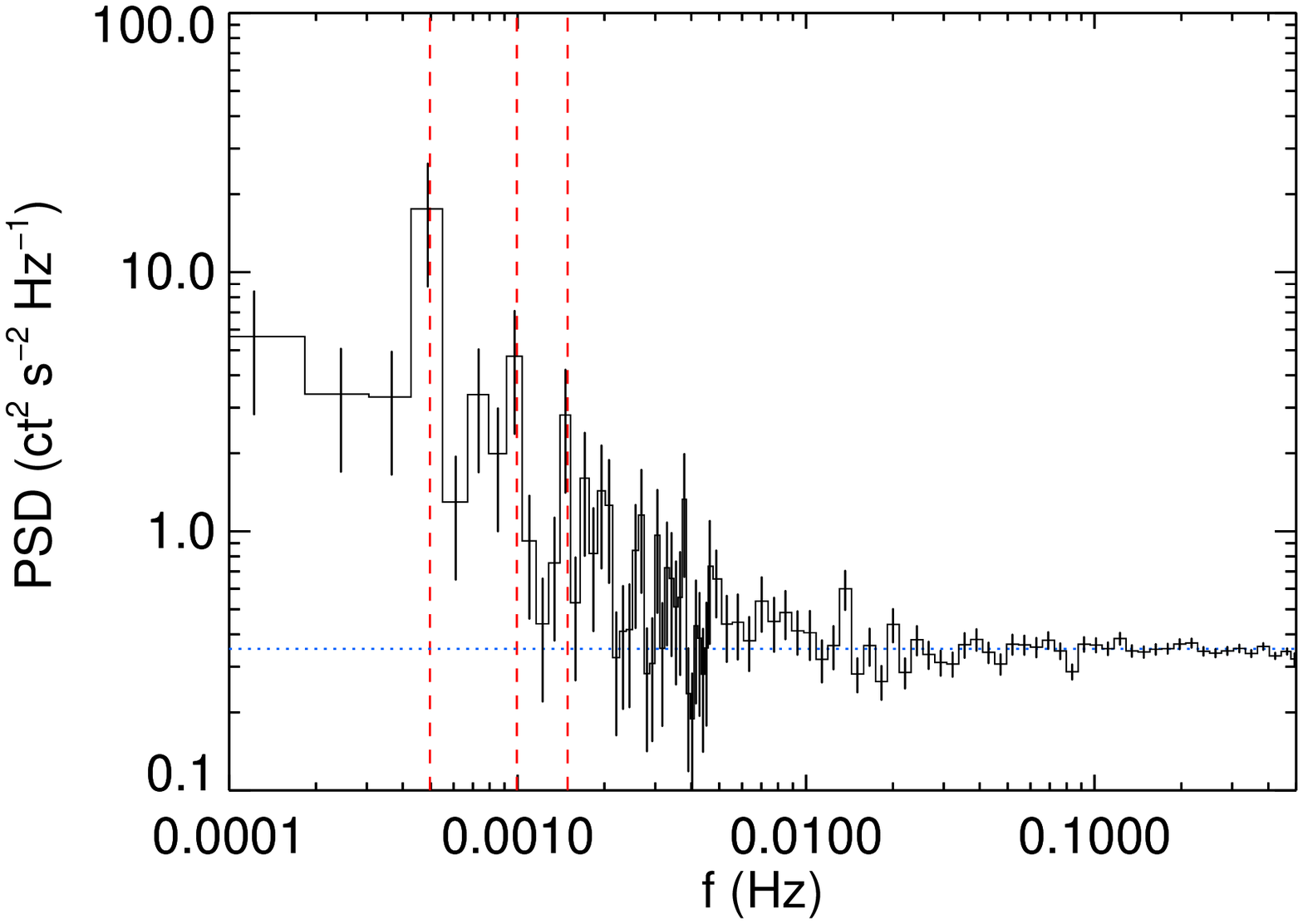}
     \includegraphics[angle=0,clip=]{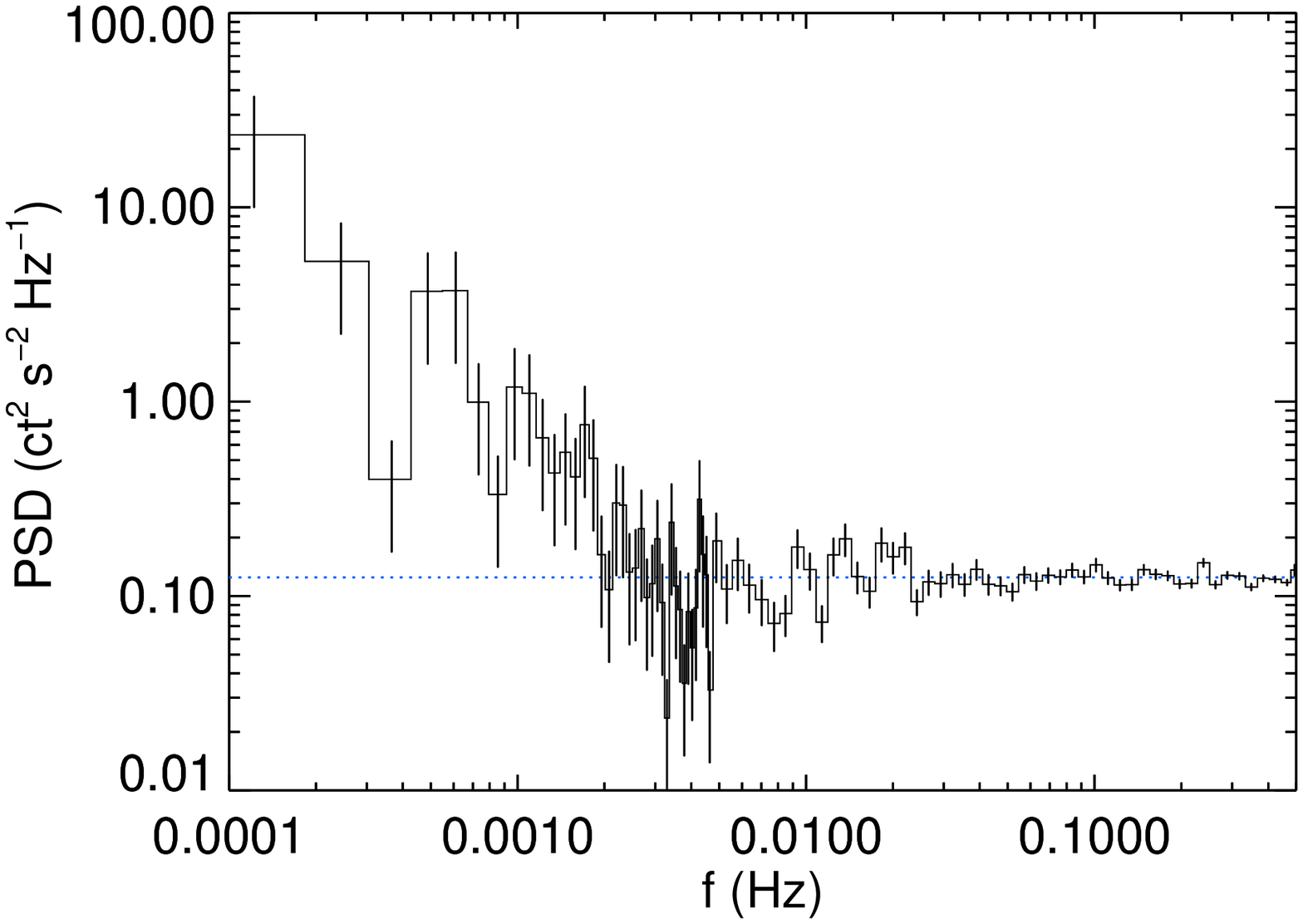}}
  \caption{Power density spectra of \canda (left) and \candb (observation April 2012, right). 
           For \canda a periodic behaviour is evident at $\sim$2013\,s seen together with two harmonics (red vertical lines).
  } 
  \label{fig:psd}
\end{figure*}

\section{Results}
\label{results}

In the following subsections, we present our results on the X-ray position, spectral and temporal properties of the two HMXB systems as well as report on the optical properties of their donor stars. 
\subsection{X-ray positions}
\label{sec:xray_pos}

X-ray positions were determined by performing a maximum-likelihood source detection analysis on the \xmm/EPIC images.
Fifteen images were created from the three EPIC cameras in five energy bands: $1\rightarrow(0.2-0.5)$ keV, $2\rightarrow(0.5-1.0)$ keV, $3\rightarrow(1.0-2.0)$ keV, $4\rightarrow(2.0-4.5)$ keV, $5\rightarrow(4.5-12.0)$ keV \citep{2009A&A...493..339W,2013A&A...558A...3S}.
Source detection was performed simultaneously on all the images using the SAS task {\tt edetect\_chain}. 
Astrometric boresight corrections were performed using a catalogue of background AGN with known redshifts or selected using ALLWISE mid-infrared colour selection criteria \citep{2012MNRAS.426.3271M,2015ApJS..221...12S} with the task {\tt eposcorr} accounting for a linear shift. The inferred shifts in right ascension and declination were applied to the attitude file of the pointings and the data reprocessed.    
For \canda we derived boresight corrected positional coordinates of 
R.A. = 05$^{\rm h}$31$^{\rm m}$08\fs33 and Dec. = $-$69\degr09\arcmin23\farcs5 (J2000),
with a $1\sigma$ statistical uncertainty of 0.1\arcsec.
From the two observations of \candb we determined an error-weighted mean of the position to 
R.A. = 05$^{\rm h}$33$^{\rm m}$20\fs87 and Dec. = $-$68\degr41\arcmin22\farcs6 (J2000)
with a $1\sigma$ statistical uncertainty of 0.55\arcsec.
The positional error is usually dominated by systematic astrometric uncertainties. Following  \citet{2013A&A...558A...3S} we added
a systematic error of 0.5\arcsec\ in quadrature. 
We note, however, that subsequent analysis of \xmm observations of the SMC with the newest SAS version has reduced the systematic uncertainty to 0.33\arcsec\ (Maitra et al., in prep).
Our analysis provides an improved X-ray position for both candidates.  
For \candb, the improved position is consistent with the one reported by \citet{2012ATel.3993....1S} confirming the initially reported counterpart \citep[i.e. blue supergiant Sk\,-68\,122;][]{1970CoTol..89.....S} as the true optical companion.

\subsection{X-ray temporal properties}
\label{sec-time_an}

Both systems exhibit strong variability during the \xmm observations (see Fig. \ref{fig:601_LC}). 
\canda shows periodic bursts every $\sim$2000\,s.
In order to compute the X-ray variability from the \xmm/EPIC light-curve we compared the highest count rate with the $5\%$ percentile of the count rate distribution. We infer an X-ray variability factor of about 10. 
\candb remained at a lower luminosity level ($\sim$0.035 counts\,s$^{-1}$) for the first $\sim$22\,ks of the exposure (obsid: 0690743801). During the last $\sim$8\,ks the light curve shows three consecutive bursts separated by $\sim$2200\,s reaching a maximum count rate of $\sim$0.45 counts\,s$^{-1}$. 
By comparing the highest count rate recorded by \xmm/EPIC with the lower $5\%$ percentile of the total X-ray light curve, we find an X-ray variability factor of at least 30.

To further investigate the X-ray variability we used the fast Fourier transform \citep{1988tns..conf...27V} and constructed the power density spectra for the two systems (see Fig. \ref{fig:psd}). The power spectra are normalized such that their integral over a range of frequencies gives the square of the root mean square (RMS). The power density spectra were grouped using a geometrical rebinning \citep{1993MNRAS.261..612P}. The spectra of both systems are dominated by white noise at high frequencies ($\gtrsim 10-30$~mHz), while they can be well described by a power-law at lower frequencies. 
The power spectrum of \canda shows additional features at $0.5$~mHz, 1~mHz, and 1.5~mHz that are related to the pulsar's spin-period and its harmonics.
In order to estimate the significance of these features and better determine the period and its uncertainty we used an epoch folding technique \citep[EF:][]{1990MNRAS.244...93D,1996A&AS..117..197L}. 
We used the combined EPIC-pn and EPIC-MOS barycentric corrected arrival times to search for a periodic signal.
We applied an EF method, while implementing the Rayleigh $Z^2_2$ normalization \citep{1983A&A...128..245B}.

\begin{figure}
  \resizebox{\hsize}{!}{\includegraphics[angle=0,clip=]{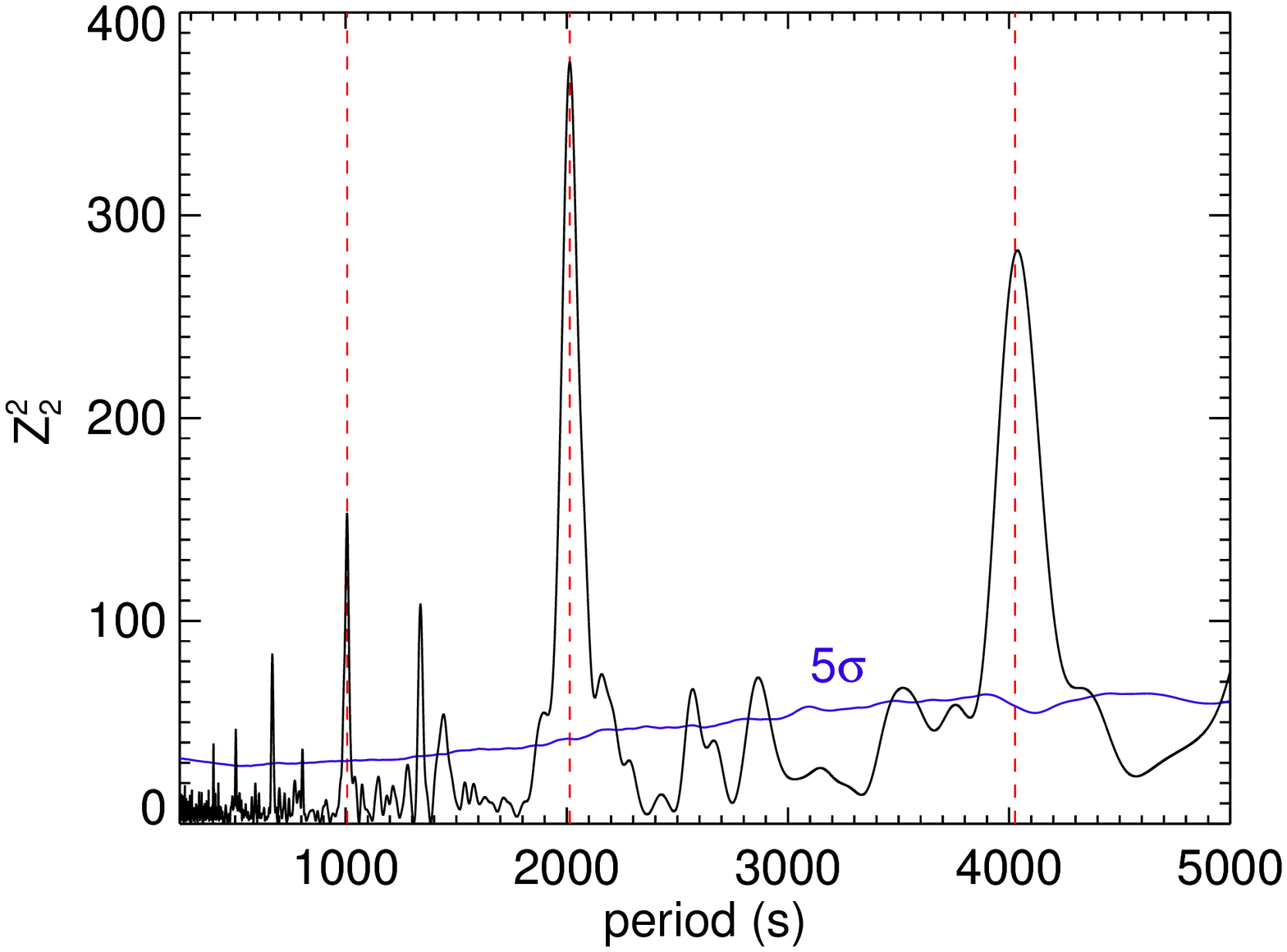}}
  \caption{Results of period search for \canda. Epoch folding (EF) method (Rayleigh $Z^2_2$ test) was applied to the combined \xmm/EPIC events (obsid:0690744601). 
  The significance of the derived period was estimated by applying the EF method on simulated light curves \citep{1995A&A...300..707T,2013MNRAS.433..907E}.}
   
  \label{fig:EF601}
\end{figure}

\begin{figure}
  \resizebox{\hsize}{!}{\includegraphics[angle=0,clip=]{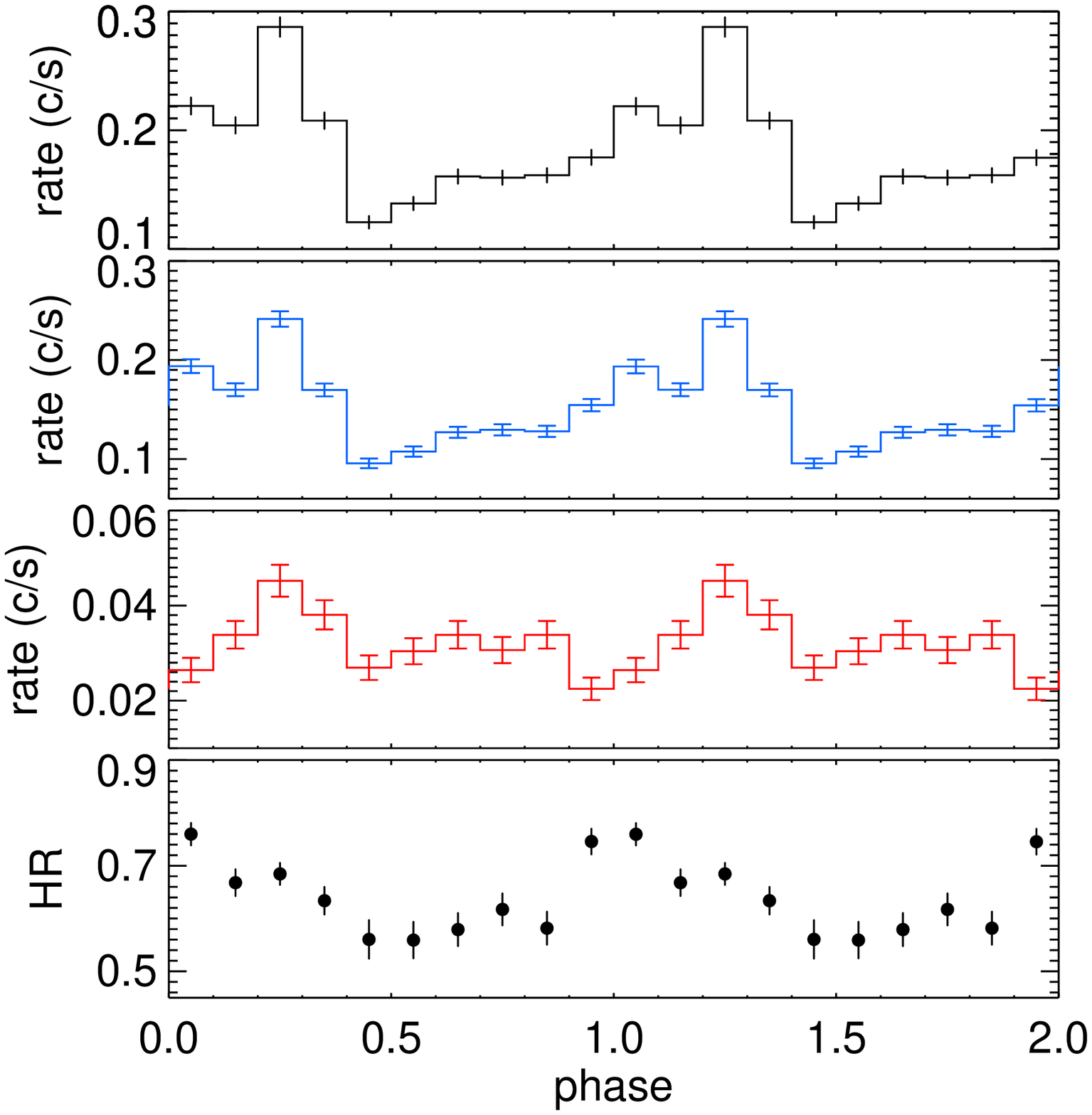}}
  \caption{From top to bottom we present the phase folded pulse profile of \canda  in the 0.3-10 keV , 2.0-10.0 keV  and 0.3-2.0 keV bands and the hardness ratio between the two sub-bands.} 
  \label{fig:pp}
\end{figure}

For \canda a clear period of 2013.5~s was found, while the periodogram shows peaks at the multiples of this period (see Fig. \ref{fig:EF601}). To estimate the uncertainty of the derived period we used two approaches. First, we analytically estimated the uncertainty to be 4-10~s using equation (14) of \citet{1986ApJ...302..757H}. Then, we performed a numerical estimation by bootstrapping the event arrival times and by performing the EF method to the bootstrapped samples. An uncertainty of $\sim$3~s was derived from the distribution of the derived periods.
The pulse-profile of the system is single peaked (Fig. \ref{fig:pp}).
We also investigated the hardness ratio (HR) evolution along the spin phase of the system by using two different energy bands (0.3-2.0, 2.0-10.0 keV). The HR is defined as $(\rm{R}_{\rm{hard}}-\rm{R}_{\rm{soft}})/(\rm{R}_{\rm{hard}}+\rm{R}_{\rm{soft}})$, with R$_{\rm hard/soft}$ denotes the background-subtracted count rates in the hard and soft energy band.
The spin phase folded light-curves of the system, for the complete energy range of \xmm as well as the soft and hard energy bands, are plotted together with the spin phase resolved HR in Fig. \ref{fig:pp}.

For \candb any periodicity search is hampered by the three late flares that dominate the statistics. Thus, we searched for periodic signal only in the first 22\,ks of the exposure. No significant period was derived from the periodicity search.

\subsection{X-ray spectral properties}
\label{sec-spec_an}

\begin{table}
\caption{Results of the X-ray spectral modelling}
\begin{center}
  \scalebox{0.9}{
 \begin{threeparttable}
\begin{tabular}{lccc}
\multicolumn{4}{c}{\canda}\\
\multicolumn{4}{l}{CASE I: constant*TBabs*TBvarabs*(powerlaw+bbodyrad)}\\
\hline
Component & Parameter & Value & units \\
\hline\noalign{\smallskip}
TBabs$^{a}$    &  nH      &  0.0618 (fixed)                 &  $10^{22} cm^{-2}$      \\ \noalign{\smallskip}  
TBvarabs    &  nH      &  2.3$\pm$0.7                 &  $10^{22} cm^{-2}$      \\ \noalign{\smallskip}  
powerlaw &  $\Gamma$ &  0.8$^{+0.8}_{-0.5}$  &     -    \\ \noalign{\smallskip}                 
         &  norm    &  1.3$^{+0.3}_{-0.5}$          &  \oergcm{-12}          \\ \noalign{\smallskip}    
bbodyrad &  $kT$ &  1.75$^{+0.5}_{-0.4}$   &     keV    \\ \noalign{\smallskip}                  
         &  norm    &  0.01$^{+0.01}_{-0.04}$          &   -         \\ \noalign{\smallskip}    
         &  R    &  0.5$^{+0.2}_{-0.1}$          &  km          \\ \noalign{\smallskip}    
\hline \noalign{\smallskip}   
\multicolumn{2}{l}{Observed Flux$^{b}$}          &     1.87$^{+0.06}_{-0.4}$               &  \oergcm{-12}         \\ \noalign{\smallskip}   
\multicolumn{2}{l}{Luminosity$^{c}$}  &    6.4$^{+0.3}_{-0.2}$           &    \oergs{35}    \\ \noalign{\smallskip}    
\hline \noalign{\smallskip}
         & $\chi^2_{\rm red}$/DOF &     0.99/252              &            \\ \noalign{\smallskip} 
\hline \noalign{\smallskip}
\hline\noalign{\smallskip}
\multicolumn{4}{l}{CASE II: constant*TBabs*TBvarabs1*(partcov*TBvarabs2)*powerlaw}\\
\hline
Component & Parameter & Value & units \\
\hline\noalign{\smallskip}
TBabs$^{a}$    &  nH      &  0.0618 (fixed)                 &  $10^{22} cm^{-2}$      \\ \noalign{\smallskip}  
TBvarabs1    &  nH      &  3.2$\pm$0.5                  &  $10^{22} cm^{-2}$      \\ \noalign{\smallskip}  
\multicolumn{4}{l}{partial covering: partcov*TBvarabs2} \\ \noalign{\smallskip} 
partcov  &  covering fraction   &  0.52$^{+0.09}_{-0.11}$          &   -         \\ \noalign{\smallskip}   
TBvarabs2&   nH &  18$^{+8.0}_{-7.0}$   &   $10^{22} cm^{-2}$     \\ \noalign{\smallskip}    
powerlaw &  $\Gamma$ &  1.30$\pm$0.20 &     -    \\ \noalign{\smallskip}                 
         &  norm    &  3.1$^{+0.6}_{-0.4}$           &  \oergcm{-12}          \\ \noalign{\smallskip}   
          
\hline \noalign{\smallskip}   
\multicolumn{2}{l}{Observed Flux$^{b}$}          &       1.87$^{+0.06}_{-0.4}$         &  \oergcm{-12}          \\ \noalign{\smallskip}   
\multicolumn{2}{l}{Luminosity$^{c}$}  &    9.4$^{+0.3}_{-0.2}$     &   \oergs{35}         \\ \noalign{\smallskip}   %
\hline \noalign{\smallskip}
         & $\chi^2_{\rm red}$/DOF &     1.00/252              &            \\ \noalign{\smallskip} 
\hline \noalign{\smallskip}
\hline\noalign{\smallskip}
\multicolumn{4}{c}{\candb}\\
\multicolumn{4}{l}{constant*TBabs*TBvarabs*powerlaw}\\
\hline
Component & Parameter & Value & units \\
\hline\noalign{\smallskip}
TBabs$^{a}$    &  nH      &  0.0609 (fixed)                 &  $10^{22} cm^{-2}$      \\ \noalign{\smallskip}  
TBvarabs    &  nH      &  0.61$^{+0.21}_{-0.18}$                  &  $10^{22} cm^{-2}$      \\ \noalign{\smallskip}  
powerlaw &  $\Gamma$ &  0.91$\pm$0.12 &     -    \\ \noalign{\smallskip}                 
         &  norm    &  0.94$\pm$0.06          &  \oergcm{-12}          \\ \noalign{\smallskip}   
\hline \noalign{\smallskip}   
\multicolumn{2}{l}{Observed Flux$^{b}$}          &     0.85$^{+0.05}_{-0.06}$           &  \oergcm{-12}          \\ \noalign{\smallskip}   
\multicolumn{2}{l}{Luminosity$^{c}$}  &    2.8$\pm$0.1     &   \oergs{35}         \\ \noalign{\smallskip}   %
\hline \noalign{\smallskip}
         & $\chi^2_{\rm red}$/DOF &     1.03/65              &            \\ \noalign{\smallskip} 
\hline \noalign{\smallskip}
\end{tabular}
\tnote{a} Contribution of Galactic foreground absorption, column density fixed to the weighted average value of \citep{1990ARA&A..28..215D} \\
\tnote{b} Absorbed flux of the fitted model in the 0.3-10.0 keV energy band. \\
\tnote{c} Unabsorbed Luminosity of the fitted model in the 0.3-10.0 keV energy band, for a distance of 50 kpc \citep{2013Natur.495...76P}. \\
\end{threeparttable}
 }
\end{center}
\label{tab:spectra}
\end{table}

X-ray spectral analysis was performed using the {\small XSPEC} fitting package, version 12.9.1 \citep{1996ASPC..101...17A}. 
We fitted the time-averaged X-ray spectra of both systems with an absorbed power-law model. 
X-ray absorption was modeled using the {\texttt tbabs}\footnote{Commonly referred to as {\texttt tbnew}, this model is now included in the latest release of {\small XSPEC}. The code is available through: \url{http://pulsar.sternwarte.uni-erlangen.de/wilms/research/tbabs/}} code \citep{2000ApJ...542..914W} with atomic cross sections adopted from \cite{1996ApJ...465..487V}.
For the modeling we used two absorption components, one to describe the Galactic foreground absorption and another to account for the column density of both the interstellar medium of the LMC and the intrinsic absorption of the sources.
For the Galactic photo-electric absorption we used a fixed column density \citep{1990ARA&A..28..215D} with abundances taken from \cite{2000ApJ...542..914W}.

\begin{figure}
    \resizebox{0.95\hsize}{!}{\includegraphics[angle=-90,clip=]{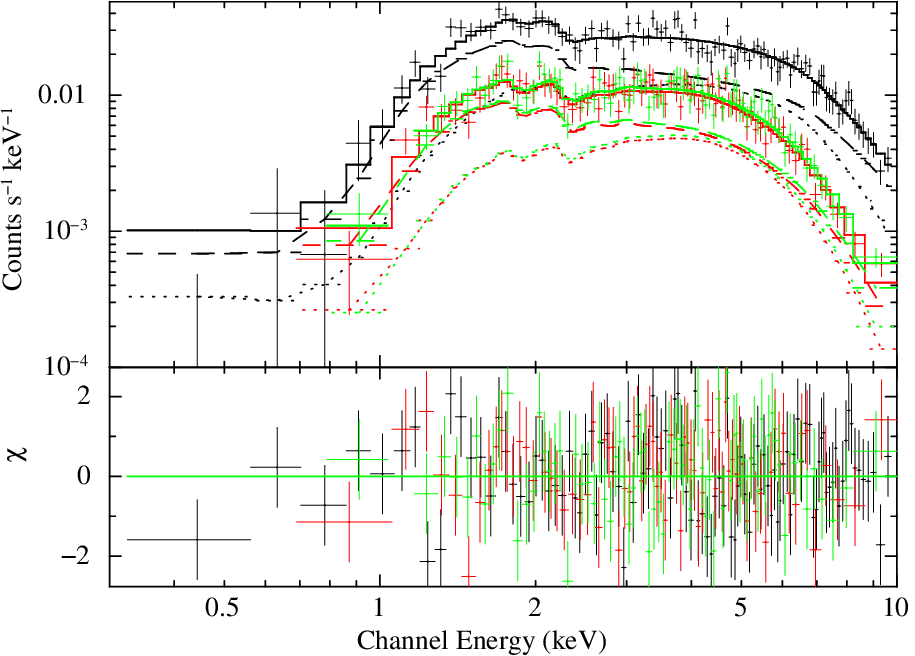}}
  \caption{EPIC spectra of \canda (black: pn, red: MOS1, green: MOS2) together with the best-fit model (solid lines) composed of 
           absorbed power-law and black-body components (CASE\,I in Table~\ref{tab:spectra}).
           The power-law (dash-dotted lines) and black-body (dotted lines) contributions are also indicated. The bottom panel 
           shows the fit residuals.}   
  \label{fig:spec_a}
\end{figure}

\begin{figure}
    \resizebox{\hsize}{!}{\includegraphics[angle=-90,clip=]{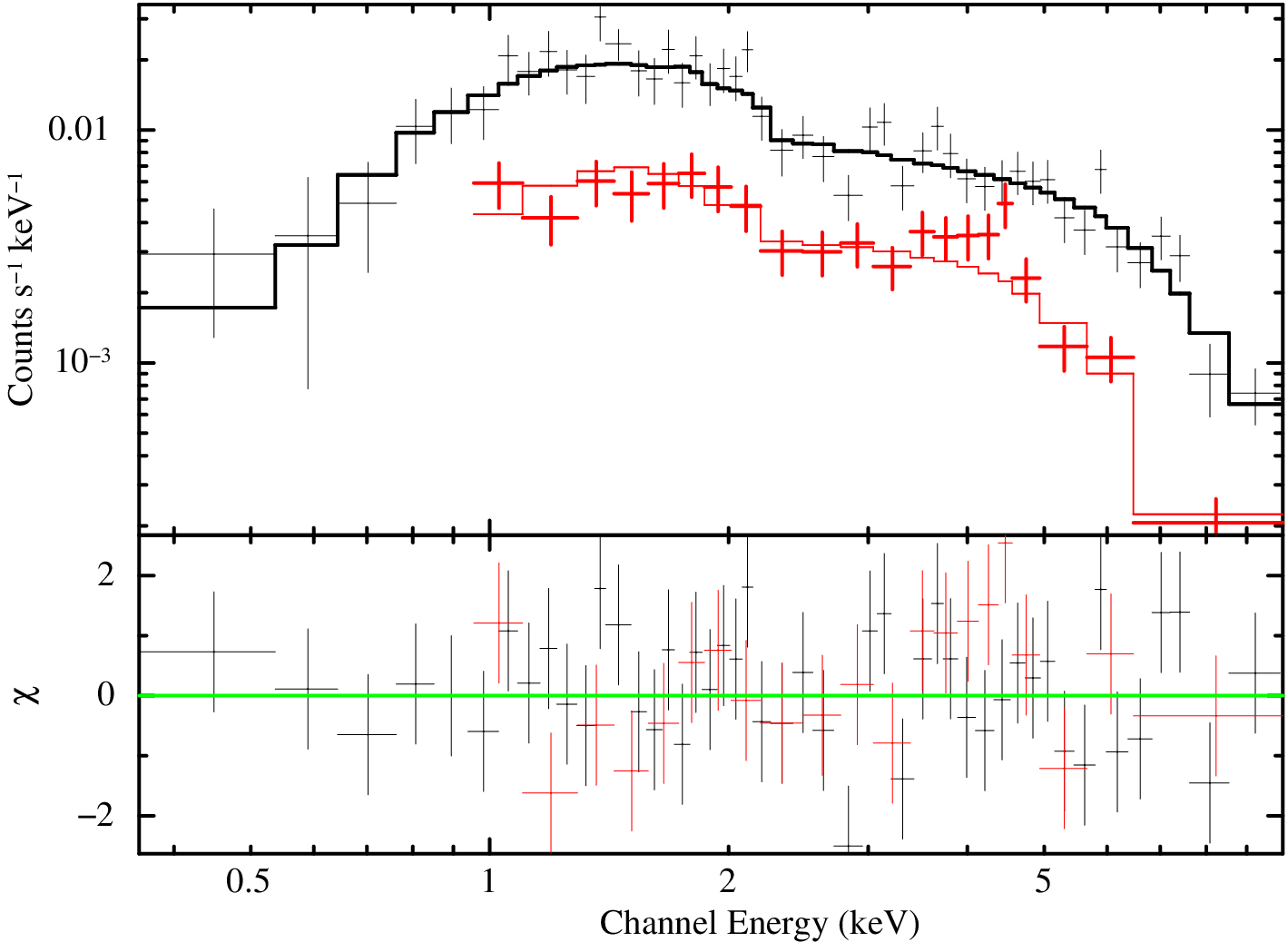}}
  \caption{EPIC spectra of \candb from April 2012 (black: pn, red: MOS2) together with the best-fit power-law model (solid lines). The bottom panel 
           shows the fit residuals.}
  \label{fig:spec_b}
\end{figure}

For \candb the X-ray spectrum was adequately described by an absorbed power-law model ($\chi^2_{\rm red}\sim1.03$). 
For \canda the absorbed power-law model yielded an acceptable goodness of fit with reduced $\chi^2\sim1.1$, but failed to describe the data in terms of residual distribution \citep{2010arXiv1012.3754A}. 
The introduction of an additional spectral component reduces the structures in the residuals. 
Such a component is commonly seen in the spectra of HMXB systems below 10 keV and is usually referred to as ``soft excess'' \citep{2004ApJ...614..881H}.
For HMXB systems observed at low luminosities this soft excess can have two different origins. 
It can be due to the NS hot spot (CASE I) and can therefore be modelled by a black-body component with typical size of 1 km radius \citep{2006A&A...455..283L,2013MNRAS.436.2054B,2013A&A...558A..74V,2017ApJ...838..133S}. 
Alternatively, it can be associated with X-ray absorption from the clumpy wind of a supergiant star (CASE II). In this case, it can be modelled by a partial-covering absorption component \citep{2009ApJ...694..344T}. Often, it is difficult to distinguish between the two cases \citep[e.g.][]{2017ApJ...841...35F}.

In the case of  \canda, both models for the soft excess provide equally good fits (CASE I: 0.99 vs CASE II: 1.01) and identical residual structure. It  is thus impossible to statistically distinguish between the two scenarios. In the case of \candb, any additional component would result in over-fitting the data and would not constrain any of the physical parameters of the system.

The X-ray spectra along with the data-to-model residual plots for both systems are presented in Fig. \ref{fig:spec_a} and \ref{fig:spec_b}.
The best fit model parameters are given in Table~\ref{tab:spectra}.
Interestingly, the intrinsic absorption of both systems is larger than that of the typical BeXRB pulsars located in the LMC \citep[e.g.][]{2014A&A...567A.129V,2016MNRAS.461.1875V,2017A&A...598A..69H}.
Moreover, by comparing the two systems, \canda is both brighter and exhibits higher intrinsic absorption.
  
From our temporal analysis (see \S\ref{sec-time_an} and Fig. \ref{fig:601_LC}), we find that \canda and \candb reached a maximum luminosity of 1.1\ergs{36} and 8.5\ergs{35} respectively during the two \xmm observations.

\subsection{Optical properties of donor stars}
\label{sec:feros}

The reduced optical spectra of the two massive stars (see Table \ref{tab:opt}) that are interpreted as the companion stars of the corresponding HMXB systems \canda and \candb, have been classified using the criteria described in \citet{2004MNRAS.353..601E,2015A&A...584A...5E}, which take into account the lower metallicity of the LMC massive stars (compared to massive Galactic stars).

\subsubsection{Spectral classification}
\begin{figure*}
  \resizebox{\hsize}{!}{\includegraphics[angle=0,clip=,bb= 14 30 550 390]{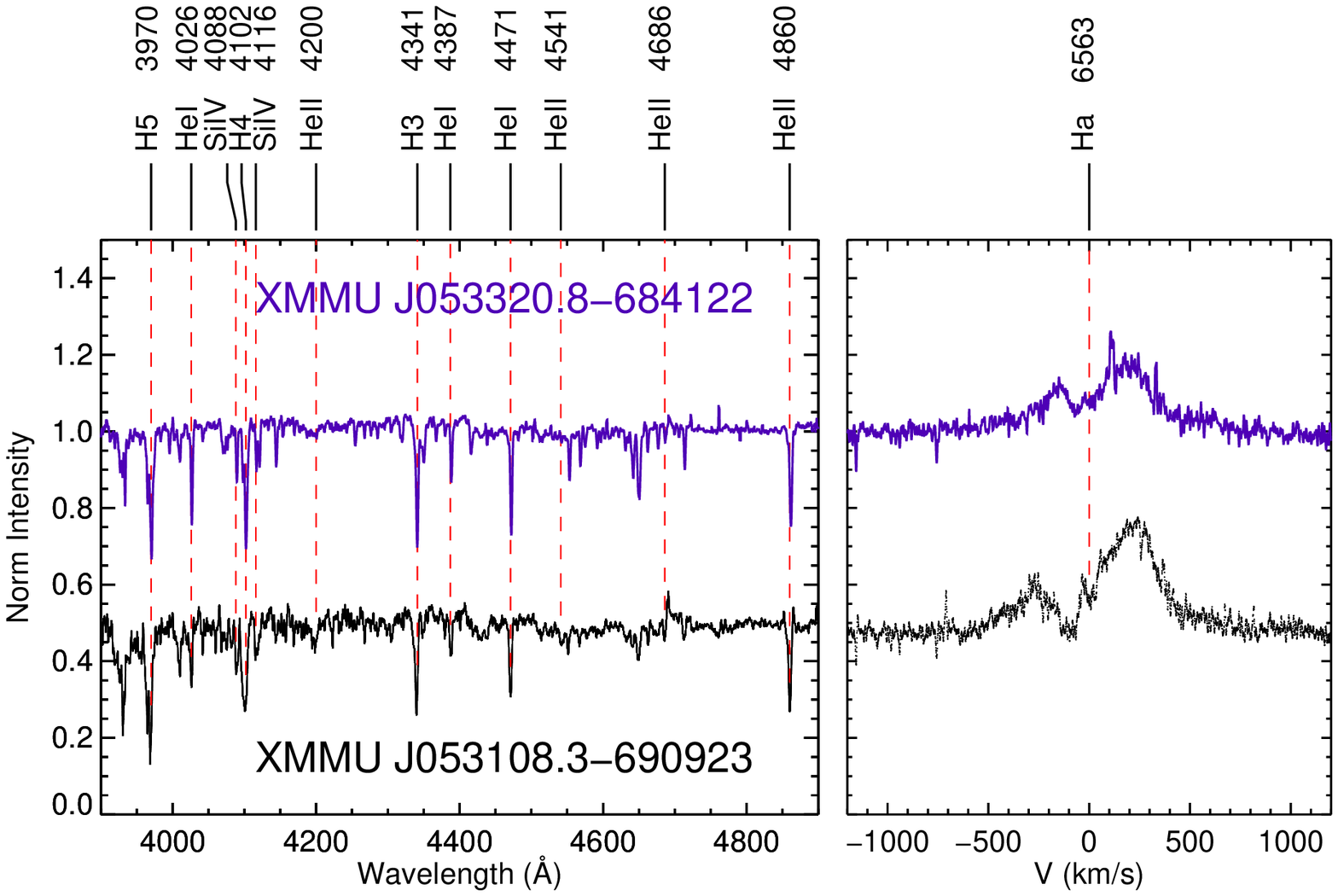}}
  \caption{FEROS spectra of \canda and \candb. Intensities have been normalised according to the underlying continuum flux and arbitrarily shifted for plotting purposes. Lines that have been used for the spectral classification are marked with vertical dashed red lines.   
  } 
  \vspace{-0.5cm}
  \label{fig:spec_fer}
\end{figure*}

The optical companion of \canda is consistent with a B0e spectral class, as absorption lines \ion{He}{II} 4686\AA~and 4541\AA~are weak but present, while the \ion{He}{II} 4200\AA~line is very weak or absent.
The optical companion of \candb is consistent with B0.5e spectral class,  as absorption lines \ion{He}{II} 4200\AA~and 4541\AA~are absent, while the \ion{He}{II} 4686\AA~line is clearly present (but weak).
Figure \ref{fig:spec_fer} shows part of the optical spectra obtained for the two stars with the main lines marked.

\subsubsection{H-alpha line profiles}
In Fig. \ref{fig:spec_fer} we show the H-alpha line region for the two stars. The narrow nebular emission lines and cosmic rays have been subtracted from the spectra.
Both profiles are double peaked. Simultaneous fits with Gaussian profiles were performed and the results are summarised in Table \ref{tab:ha}.
Usually, the profiles of supergiants of this type are expected to show P-Cygni profiles in the Balmer lines. However the absence of P-Cygni signature does not prove that a star is not a supergiant, as the profile is variable and goes through phases with no P-Cygni signature \citep[e.g. ][]{2007ApJ...656..437G,2006ESASP.604..165N}.

\subsubsection{Luminosity Classification}
The luminosity class of the systems was determined using a combination of spectroscopic and hybrid spectro-photometric criteria. 

Spectroscopically, the spectra of both \canda and \candb show clear \ion{Si}{III} lines (4553\AA, 4568\AA, 4575\AA), \ion{N}{III} 4097\AA, and \ion{O}{II} lines (4640\AA, 4643\AA, 4650\AA, 4699\AA, 4705\AA), that indicate a supergiant luminosity class \citep{2008MNRAS.388.1127H}. 
The equivalent width of the Balmer H$_{\gamma}$ line is $\simeq$1.06\AA~ for the case of \canda and $\simeq$1.17\AA~ for \candb. Both values are consistent with a supergiant luminosity class given the spectral type of the objects \citep[see e.g.][]{1993A&AS..101..599C}. 

Classification based on photometric or spectrophotometric criteria requires an estimate of the absolute magnitude of the systems, which can be derived from the apparent magnitudes of the objects given in Table \ref{tab:opt}, corrected for the distance modulus of the LMC of $\mu_0$=18.49 mag
\citep{2013Natur.495...76P} and for interstellar extinction. As an indication of the interstellar absorption in the direction of \canda and \candb, we used the values provided by \citet{2011AJ....141..158H}. By averaging the provided values around a 10\arcmin~area around the two systems we derived E(B-V) values of 0.076 and 0.086 for \canda and \candb ,  respectively. Given the spectral class of the two stars (B0-B0.5) one would expect a (B-V)$_0$ colour of $\simeq$-0.2. Comparing with the actual B-V colours from Table 2, reddenings of $\simeq$0.28~mag for \canda and $\simeq$0.11mag for \candb are inferred. For the case of \candb the two reddening estimates are consistent, while in the case of \canda, the actual reddening of the star is significantly higher than expected according to \citet{2011AJ....141..158H}, probably indicating the presence of absorption related to the system itself, or, alternatively the colour of the star may be affected by the contribution of a circumstellar disk. Adopting the first set of E(B-V) values, and applying the corresponding extinction correction, we derive, M$_V= -5.07$ and   M$_V= -6.05$ for \canda and \candb respectively.
For the second set E(B-V) estimates the corresponding absolute magnitudes of M$_V= -5.70$ and   M$_V= -6.11$ for \canda and \candb respectively, assuming that the observed B-V colour is only due to reddening and not to the contribution from a possible disk. 

Given these values for the absolute magnitudes of the two donor stars and the values for the equivalent widths of the H$_{\gamma}$ line mentioned earlier, we can apply the hybrid luminosity class criteria appropriate for Magellanic Cloud metallicity \citep[see Table 3 of][]{2004MNRAS.353..601E}. Note that the ranges given in this paper refer to a distance modulus of 19.2 (for the SMC). After taking this into account, \canda can be assigned a hybrid luminosity class of (II), while \candb is clearly of luminosity class (Ib), with the parentheses indicating that these are not truly morphological luminosity classes, but are based on both spectroscopic and photometric criteria.  

Using only photometric criteria, based on the expected range of absolute magnitudes for B0-B0.5 spectral types of different luminosity classes \citep[e.g.][]{2006MNRAS.371..185W,2017AJ....154..102U}, both \canda and \candb are consistent with luminosity class I supergiants. 
 
To summarize, all classification criteria indicate that \canda is a B0 II-Ib bright giant or supergiant and \candb is a B0.5 Ib supergiant. For \canda there is indication of either heavy local reddening, or of contribution of a circum-stellar disk component \citep[even present in supergiant systems;][]{2017arXiv171002585M}.


\begin{table}
 \centering 
 \caption {H-alpha profile characteristics}
 	\label{tab:ha}
 \begin{tabular}{lccc}
\hline
\multicolumn{4}{l}{ \candb}\\
& $\lambda_{\rm central,1}$ (\AA) & FWHM (\AA) & EW \\
\hline
Red & 6558.8$\pm$0.1 &  4.1$\pm$0.4 & -0.44$\pm$0.03\\
Blue & 6566.9$\pm$0.1 &  7.8$\pm$0.2  & -1.35$\pm$0.04\\
\hline
\hline \noalign{\smallskip}
\multicolumn{4}{l}{ \canda}\\
& $\lambda_{\rm central,1}$ (\AA) & FWHM (\AA) & EW \\
\hline
Red & 6556.54$\pm$0.06 & 4.8$\pm$0.2 & -0.51$\pm$0.02 \\
Blue & 6567.50$\pm$0.03 & 7.1$\pm$0.1 & -2.06$\pm$0.02 \\
\hline 

\end{tabular}
\end{table}

\begin{figure}
    \resizebox{\hsize}{!}{\includegraphics[angle=0,clip=]{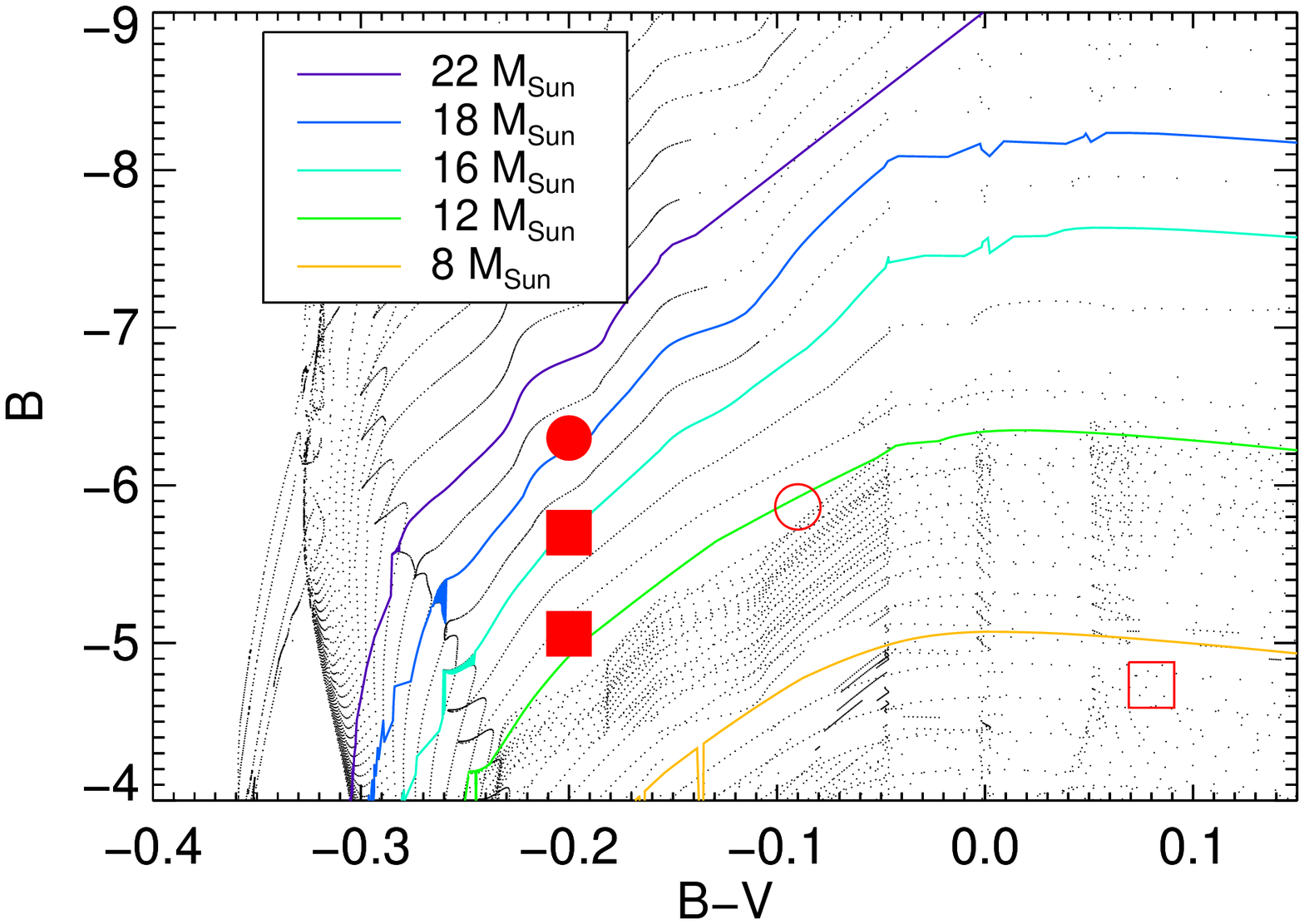}}
  \caption{Colour-magnitude diagram of the MESA evolutionary models for the LMC metallicity \citep{2016ApJ...823..102C,2016ApJS..222....8D}. Stellar tracks for different mass-ranges are overplotted with coloured solid lines. The position of \canda (squares) and \candb (circles) in the diagram are marked using both their corrected (filled symbols) and uncorrected (open symbols) colour and magnitude values. For \canda the two corrected values correspond to the case of local extinction (brighter) or disk contribution (see text for details).}
  \label{fig:mist}
\end{figure}

\subsubsection{Masses of donor stars}
For both systems we can use their combined spectro-photometric properties to estimate the masses of the donor stars.
For this purpose we can use the Modules for Experiments in Stellar Astrophysics \citep[MESA;][]{2011ApJS..192....3P} Isochrones and Stellar Tracks \citep[MIST][]{2016ApJ...823..102C,2016ApJS..222....8D}. 
We compared the photometric properties of the systems with the MIST curves produced for the LMC metallicity taking into account the effect of rotation on stellar models. This is presented in Fig. \ref{fig:mist}. 
In both systems, the $(B-V)_0$ colour of the star (i.e its temperature) is better determined by its spectral class than by its photometric properties. 
The photometric color is affected by the uncertainties in interstellar and circumstellar reddening and the possible contribution from a circum-stellar disk.
Therefore, we used $B-V\simeq-0.2$ as the de-redenned colour of the donor.
Regarding the absolute magnitude, we use the values calculated earlier, with the clarification that in the case of \canda the absolute magnitude can have an uncertainty of the order of 0.5 mag in B. This is a result of the different interpretations that are possible for the observed large B-V value. If it is caused by high local or circumstellar absorption, then the corresponding extinction needs to be taken into account in the derivation of $M_B$ (leading to the brighter value of $M_B$).  This interpretation is supported by the high absorption of the X-ray spectrum.  If, on the other hand, the high B-V value is due to a disk component, which is expected to contribute mostly to the red and infrared, $M_B$ is not expected to be affected appreciably. In this case, we only correct for the known interstellar extinction in the region from \citet{2011AJ....141..158H}. Interestingly,  the ratio between E(B-V) and the equivalent width of the H$_{\alpha}$ line is consistent with the expected relation for BeXRB systems \citep{2012A&A...539A.114R}, thus supporting this interpretation.

Taking into account all these uncertainties, and according to Fig. \ref{fig:mist}, we estimate that the mass of \canda is in the range 12-18 M$_{\sun}$ and of \candb around 18-22 M$_{\sun}$. We note that the upper limits on these ranges are mainly driven from our assumption on the expected color of the donor star that was assumed $B-V\simeq-0.2$, but could be as low as -0.25.

\section{Discussion}
 \label{discussion}

We have presented a detailed study of the X-ray spectral and temporal properties of two newly discovered HMXB systems in the direction of the LMC.
Below we summarize the results of our study and discuss the properties of the systems in the context of quasi-spherical accretion in the subsonic regime. 

\subsection{\canda: a $\sim$2013 s pulsar with a bright giant (or supergiant Ib) companion}

Timing analysis of the \xmm data revealed the presence of coherent pulsations with a period of $\sim$2013~s. The phase resolved hardness ratio of the system (see Fig. \ref{fig:pp}) 
reveals a week correlation between its luminosity and hardness ratio (harder when brighter), which is in agreement with the general trend of accreting XRB pulsars in the MCs \citep[e.g.][]{2014A&A...567A.129V,2014MNRAS.444.3571S,koliopanos17}.
The optical companion of the system was identified as a B0e II-Ib with a mass in the range of 12-18 M\sun~(see \S~\ref{sec:feros}).

From all the available X-ray observations we can set a lower limit on the long-term (i.e. not due to pulsations) X-ray variability of the system (factor of 8; see Table \ref{tab:log}). 
This is relatively low compared to pulsating HMXBs located in the MCs \citep[for SMC systems]{2016A&A...586A..81H}. 
The small long-term variability of the system can be a result of the nature of the donor star (i.e. supergiant).
BeXRBs can exhibit variability in their X-ray light curves on timescales of months to years that can be connected to the activity of the donor star and changes of its decretion disk \citep[e.g.][]{2010MNRAS.401...55R}.
Very long optical variations related to modulations of the Be disk size is present in most of the SMC pulsars \citep{2011MNRAS.413.1600R}. 
Similarly, analysis of optical monitoring data of SMC BeXRB pulsars obtained by OGLE\footnote{OGLE: Optical Gravitational Lensing Experiment \citep{2015AcA....65....1U}}, have revealed optical periodicities that can be interpreted as orbital period or non-radial pulsations of the Be star \citep{2013MNRAS.431..252S,2017MNRAS.467.1526M}.
In many cases, orbital periods of individual BeXRB systems in both the LMC and SMC have been inferred from their OGLE optical light-curves\footnote{The OGLE-IV I band photometric data are publicly available online through the OGLE real time monitoring of X-ray variables web page \citep{2008AcA....58..187U}: \url{http://ogle.astrouw.edu.pl/ogle4/xrom/xrom.html}} \citep[e.g.][]{2016MNRAS.461.1875V,2017MNRAS.470.4354V,2014MNRAS.444.3571S,2017MNRAS.466.1149B}.
It is therefore atypical for BeXRB systems not to exhibit some sort of optical variability due to the presence of a disk around the massive star.
We analysed the available OGLE data of \canda (Udalski,~A.: private communication) using the Lomb-Scargle periodogram \citep{1982ApJ...263..835S,1986ApJ...302..757H}.
The OGLE-IV light curve of \canda reveal no long-term modulation or periodic signal that could be interpreted as orbital period \citep[see also][]{naomi17conf}.
Due to the above, we interpret the measured pulsations as the NS spin period, and classify the system as a SgXRB pulsar, the 2$^{\rm nd}$ known in the LMC. In this scenario the long-term X-ray variability of the system is easily explained in terms of orbital modulation due to an eccentric orbit \citep[e.g.][]{1991ApJ...376..245H}.

The X-ray spectrum of the system can be satisfactorily described by (i) an absorbed power law plus black-body component or by (ii) a partial-covered absorbed power law (see \S\ref{sec-spec_an}). In the first scenario,  the black-body component has a temperature of $\sim$1.75 keV and a size of $\sim$0.5~km, and can be explained by the presence of a hot spot at the NS magnetic pole. 
In the second interpretation, the high X-ray absorption can be associated with the presence of a clumpy wind originating from the supergiant star. Another physical alternative for explaining the presence of a partial-covering absorber is that the NS is covered by a dense shell of material or an accretion curtain. In this scenario, any changes in the HR with pulse profile may be explained by variations in the covering fraction with pulse phase, if the covering material is locked with the NS rotation period. To investigate this possibility, we 
used the best-fit model to create simulated spectra for different values of the covering fraction. The minimum observed HR ($\sim$0.57 at phase $\sim$0.5) can be reproduced by a covering factor of 0.33, whereas the maximum observed HR ($\sim$0.74 at phase $\sim$1) requires a covering factor of 0.8. Nevertheless, this interpretation makes it difficult to explain such high absorptions from ionized gas as it would require very high densities and the resulting optical depth would probe much higher accretion rates than the one inferred from the observed X-ray luminosity.    

\subsection{\candb: A probable SFXT in the LMC}

Timing analysis of the X-ray events collected by \xmm (obsid: 0690743801)  did not reveal the presence of any statistically significant periodic signal. 
During the first 22 ks of the \xmm observation, the X-ray luminosity of the system shows small variations around a mean value of $L_{\rm X,low} \sim1.4$~\ergs{35}. Towards the end of the observation the system's luminosity increased rapidly (within $\sim$0.5 ks) by a factor of $\sim$13, while three subsequent flares appear to occur every $\sim$2200 s (see Fig. \ref{fig:601_LC}). 
We propose that this indicates the onset of fast bright X-ray flares which have been detected in HMXB systems with supergiant companions \citep{2006ESASP.604..165N}. 
This is also consistent with the quasi-periodic flaring activity observed in other galactic SFXT \citep[e.g. IGR J11215-5952 \& IGR J16418-4532:][]{2017ApJ...838..133S,2012MNRAS.420..554S}. 

Following this scenario the low X-ray luminosity of \candb prior to the three flares may be the result of quasi-spherical settling accretion in a wind-fed HMXB (\citealt{2012MNRAS.420..216S}; see also \citealt{2017arXiv170203393S} for a recent review). 
The optical counterpart of \candb, which has been identified as a B0.5e Ib star with a mass of $18-22 M_{\sun}$ (see Fig.~\ref{fig:mist}), is also compatible with the classification of the system as an SFXT.

The X-ray spectrum of the system can be described well by an absorbed power-law model, but due to low photon statistics we were not able to test more complex models. Nevertheless, the high value of intrinsic column density that is required to model the system's X-ray spectrum is in favour of a dense environment expected in a wind-fed accreting system. 
It is typical for the X-ray spectrum of wind-fed systems to exhibit large intrinsic absorption.
Obscured HMXBs are naturally explained by a compact object orbiting inside a cocoon of dust and/or cold gas formed by a strong and clumpy stellar wind of the companion \citep{2008ChJAS...8..197C,2008arXiv0809.1076C,2012MmSAI..83..251B}.
Typical SgXRBs show larger persistent luminosities and higher absorption (\nh$>$\ohcm{22}) indicating a stronger wind, while SFXTs are known to show lower intrinsic absorption values (\ohcm{21}$<$\nh$<$\ohcm{22}). The latter is consistent with the scenario that \candb is most probably an SFXT system.

\subsubsection{Quasi-spherical accretion onto an NS}
By adopting the scenario of quasi-spherical accretion onto a slowly rotating NS, we provide rough estimates for the physical properties of \candb.

In the scenario of quasi-spherical accretion, the stellar wind material that is being gravitationally captured by the orbiting NS can either fall supersonically towards the NS magnetosphere (supersonic regime) or it can form a quasi-static shell of hot plasma above it (subsonic regime) \citep{2012MNRAS.420..216S}.  The latter is relevant for systems with $L_{\rm X}\lesssim 4\times 10^{36}$~erg s$^{-1}$, whereas the supersonic regime settles in for higher X-ray luminosities. Thus, the quasi-spherical accretion model is applicable to both persistent and SFXT systems with slowly rotating pulsars.

As the X-ray luminosity of \candb never exceeds 4\ergs{36} for the duration of the \xmm observation (obsid: 0690743801), it is reasonable to assume that accretion takes place in the subsonic regime.
The entry of accreting matter through the magnetosphere determines the X-ray luminosity from the NS\footnote{We assume  $L_{\rm X}=\epsilon \dot{M}c^2$, where $\epsilon=0.1$ is the accretion efficiency.} and is being regulated by the plasma cooling processes. These lead effectively to a lower radial plasma velocity ($v_{\rm r}$) than the free-fall velocity at a given radius $r$ ($v_{\rm ff}=\sqrt{2GM_{\rm NS}/r}$) and can be incorporated into the dimensionless factor $f(v)=v_{\rm r}/v_{\rm ff} \lesssim 1$. For $L_{\rm X} \gtrsim 10^{35}$~erg s$^{-1}$, which is relevant to \candb,  Compton cooling dominates over the radiative cooling due to free-free emission and $f_{\rm C}$ is given by \citep{2017arXiv170203393S}:
\eqb 
f_{\rm C}(v) \approx 0.22 \, \zeta^{7/11}\dot{M}_{16}^{4/11}\mu_{30}^{-1/11},
\label{eq:fc}
\eqe 
where $\zeta\lesssim 1$, $\dot{M}=10^{16} \dot{M}_{16}$~g s$^{-1}$ is the accretion rate inferred from the X-ray luminosity, and $\mu=10^{30}\mu_{30}$ G cm$^3$ is the NS magnetic moment. The X-ray luminosity of the non-flaring state can be then derived \citep{2014MNRAS.442.2325S}:
\eqb 
& &\frac{L_{\rm X, low}}{5\times 10^{35} {\rm erg \ s}^{-1}} \simeq  \\ \nonumber 
& & f_{\rm C}(v) \left(\frac{M}{10 M_{\odot}}\right)^{s-2/3} \left(\frac{v_{\infty}}{10^{3} \ {\rm km \ s^{-1}} }\right)^{-1}\left(\frac{v_{\rm w}}{500 \ {\rm km \ s^{-1}}}\right)^{-4}\left(\frac{P_{\rm orb}}{10 \ {\rm d}} \right)^{-4/3},
\label{eq:Lx}
\eqe 
where $s=2.76$ is the power law index of the phenomenological mass-luminosity relation for massive stars \citep[see][]{2007AstL...33..251V}, $v_{\infty}$ is the terminal wind velocity of massive stars (typically 1000 km s$^{-1}$), $v_{\rm w}$ is the wind velocity at the NS Bondi radius, and $P_{\rm b}$ is the orbital period. Hence, we adopt $M=18 M_{\odot}$ as an reference value (see Fig. \ref{fig:mist}). 

An estimate of $v_{\rm w}$ can be obtained, if one matches the mean duration of the observed X-ray flares ($t_{\rm fl} \sim 2.2\times 10^3$~s) with the free-fall timescale from the outer radius of the plasma shell (i.e., the Bondi radius):
\eqb 
v_{\rm w} = \left(\frac{2GM_{\rm NS}}{t_{\rm fl}}\right)^{1/3} =  570\ \left(\frac{M_{\rm NS}}{1.4 \,  M_{\odot}}\right)^{1/3}\left(\frac{t_{\rm fl}}{10^3\, {\rm s}}\right)^{-1/3} {\rm km \ s}^{-1},
\label{eq:vw}
\eqe 
By substituting  eqs.~(\ref{eq:fc}), (\ref{eq:vw}) and the measured mean X-ray luminosity (i.e, $L_{\rm X, low} \simeq 1.4\times 10^{35}$~erg s$^{-1}$) into eq.~(\ref{eq:Lx}), we find that  $P_{\rm orb}\sim 10$~d. Interestingly, the derived orbital period is similar to those of supergiant and SFXT systems \citep[e.g.][]{2012A&A...539A..21D}.   

The torques acting on the quasi-static plasma shell above the magnetosphere can either spin-up or spin-down the NS. Because of the short relaxation time ($\sim$$10^4$~yr -- see also \citealt{2017MNRAS.469.3056S}), it is reasonable to assume that the NS is at an equilibrium state, where its spin period remains constant with time and is given by \citep{2012MNRAS.420..216S,2017arXiv170203393S}:
\eqb
P_{\rm s,eq} \approx 1000 \ \mu_{30}^{12/11} \dot{M}_{16}^{-4/11} \left( \frac{P_{\rm d}}{10 \ {\rm d}}\right) \left(\frac{v_{\rm w}}{10^3 \ {\rm km \ s^{-1}}}\right)^4 {\rm s}
\label{eq:peq}
\eqe 
Using the estimated values for the wind velocity and the orbital period, we find $P_{\rm eq} \sim 200$~s for $\mu_{30}=1$. Below we list other parameters of the shell that we can estimate:
\begin{itemize}
 \item Bondi radius: $R_{\rm B}\simeq 2 G M_{\rm NS}/v_{\rm w}^2 \sim 10^{11}$~cm
 \item Alfv{\'e}n radius \citep{2012MNRAS.420..216S,2017arXiv170203393S}: 
 \eqb 
 R_{\rm A}\simeq \left( \frac{30  \Gamma}{\Gamma -1} \frac{f_{\rm C}(v) \mu^2}{\dot{M} \sqrt{2GM_{\rm NS}}}\right)^{2/7} \sim 2\times 10^9 {\rm cm},
 \eqe 
 where $\Gamma=5/3$ is the adiabatic index.
 \item Temperature of the plasma at  Alfv{\'e}n  radius: 
 \eqb 
 T\left(R_{\rm A}\right) \approx \frac{\Gamma-1}{\Gamma}\frac{GM_{\rm NS}\bar{\mu}}{\mathcal{R} R_{\rm A}} \sim 2.5\times 10^8 \, {\rm K} 
 \eqe 
 where $\bar{\mu}\simeq 0.60$ for ionized gas for the LMC chemical composition \citep{2002A&A...396...53R} and $\mathcal{R}$ is the gas constant. The derived temperature is sufficiently low to allow the entry of matter through the magnetosphere via the interchange instability \citep[see][]{2012MNRAS.420..216S}.  
 \item Mass of the plasma shell \citep{2014MNRAS.442.2325S}:
 \eqb 
 M_{\rm sh}\approx \frac{2}{3}\frac{\dot{M}}{f_{\rm C}(v)}t_{\rm ff}(R_{\rm B}) \simeq 2\times 10^{19} \ {\rm g}.
 \eqe 
 \item Plasma density at Alfv{\'e}n radius \citep{2012MNRAS.420..216S}: 
 \eqb 
 \rho\left(R_{\rm A}\right)\approx \frac{3 M_{\rm sh}}{8\pi R_{\rm A}^{3/3} \left(R_{\rm B}^{3/2}-R_{\rm A}^{3/2}\right)} \simeq 8\times 10^{-13} \ {\rm g \ cm^{-3}},
 \eqe
 where a density profile of $\rho (R) \propto R^{-3/2}$ was assumed.
 \end{itemize}

\begin{figure}
    \resizebox{\hsize}{!}{\includegraphics[angle=0,clip=]{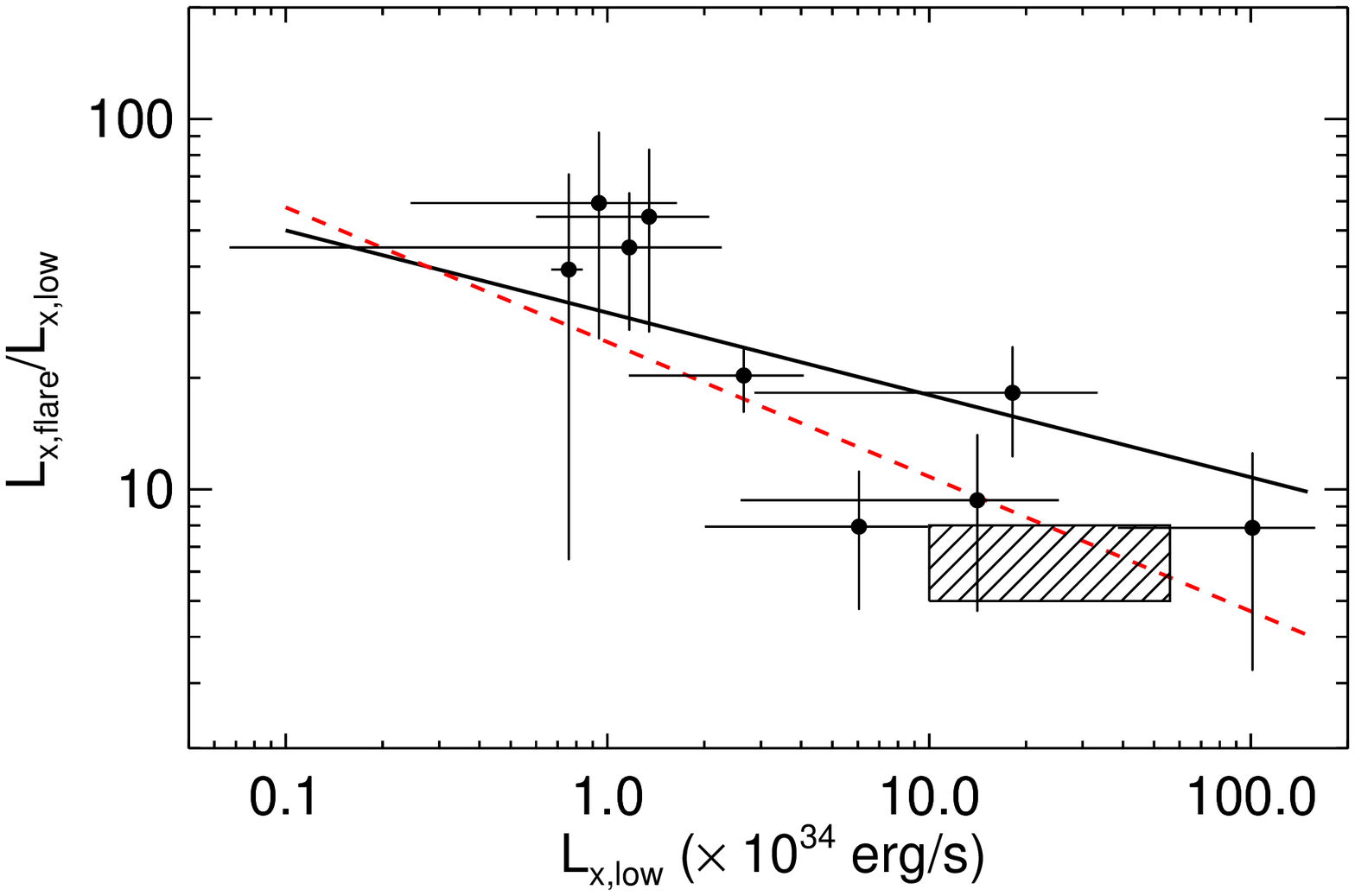}}
  \caption{Comparison of the mean dynamic flare range of other known SFXT systems. Values for the known systems (black crosses) refer to the INTEGRAL/IBIS energy range (17-50 keV) and were obtained from \citet{2014MNRAS.439.3439P}. For \candb~(striped band) we used the \xmm derived rates in the 0.3-10 keV energy range and extrapolated to the 17-50 keV band by assuming a power law with photon index 0.92 with different cutoff energies (10 \& 50 keV).  Overplotted are the expected dependencies of $L_{\rm X, flare}/L_{\rm X,low}$ on the non-flaring luminosity in two regimes of plasma cooling in the shell: radiative cooling (black dotted line)  \citep[see][]{2014MNRAS.442.2325S} or Compton cooling (red dashed line), as described in the text.}
  \label{fig:sfxr_flare}
\end{figure}

In the Compton cooling regime, the ratio of the flaring  to the non-flaring luminosity can be estimated as $L_{\rm X, flare}/L_{\rm X, low} \approx 1/f_{\rm C}(v)\propto L_{\rm X, low}^{-4/11}$.  Using the mean low luminosity of \candb in the 0.3-10 keV and equation (\ref{eq:fc}), we find  $L_{\rm X, flare}/L_{\rm X, low} \approx 7$. 
Following this, we made a comparison of our results with those derived for other known SFXT systems. 
Recently, \cite{2014MNRAS.442.2325S} computed the mean dynamic range of INTEGRAL detected flares from SFXT systems by using long-term INTEGRAL archival data in hard X-rays (17-50~keV) \citep{2014MNRAS.439.3439P}. This is illustrated in Fig.~\ref{fig:sfxr_flare} (black crosses).
\candb is well modeled by an absorbed power-law in the 0.3-10 keV band.
The X-ray spectrum of accreting NS exhibits a cutoff above 10~keV, thus it is not straight forward to extrapolate the luminosity of accreting NS in different energy bands. Nevertheless, we can estimate an expected range for its hard X-ray luminosity assuming an upper (i.e. 50~keV) and lower (i.e. 10~keV) limit for the cutoff energy.
If we assume a cutoff power law with photon index 0.92 (see Table~\ref{tab:spectra}) and variable cutoff energy,
the non-flaring luminosity of \candb in the 17-50 keV band ranges between $\sim$1 and 4 times its luminosity in the 0.3-10.0 keV band.
The variability of the system in the soft and hard X-rays is expected to be the same, assuming no significant change in its spectral properties during the flares.
The location of \candb in the $L_{\rm X,low}$ vs $L_{\rm X,flare}/L_{\rm X,low}$ diagram, shown as a striped band in Fig.~\ref{fig:sfxr_flare}, is in agreement with the theoretical predictions (dashed and dotted lines). We caution the reader that our estimates for the time-averaged luminosities are limited to a single \xmm observation.

In principle, similar estimates can be performed for \canda by assuming that the observed spin-period corresponds to the equilibrium one (see eq.~(\ref{eq:peq})).
However, the estimates will suffer from large uncertainties due to the following reasons: No fast bright flares have been detected, thus no flare duration and amplitude can be determined.
Subsequently, no wind velocity or orbital period can be estimated.
We note that $P_{\rm s, eq}$ has a strong dependence on the wind velocity \citep[see also ][]{2014MNRAS.442.2325S}. 
Finally, the nature of the donor star is less certain than in the \candb system.

\section{Conclusions}

We present a detailed analysis of two SgXRBs in the LMC, raising the number of LMC SgXRB systems to five. We study the X-ray timing and spectral properties of the compact objects as well as the optical properties of the donor stars. 

The first system \canda, is a 2013.5 s pulsar having a supergiant (or bright giant) companion, this is the 2$^{\rm nd}$ confirmed SgXRB pulsar in the LMC. Its X-ray spectrum is consistent with an absorbed power-law with a soft excess which can be either interpreted to originate from the hot spots of the NS, or as the result of partial absorption by the clumpy stellar wind of the giant companion.  We further find indications of the variation of the hardness ratio with the pulse phase. The optical companion is identified as B0e II-Ib with mass in the range of 12-18 M$\odot$.

The second system \candb,  is the first identified SFXT candidate in the LMC. The X-ray light curve, obtained by \xmm, displays a long period of quiescence followed by three subsequent flares of high dynamic range. The spectrum is consistent with an absorbed power-law with a moderate value of absorption column density. The optical companion is identified as B0.5e 1b with a mass of 18 M$\odot$. 
The above properties are indicative of the SFXT nature of the system.

\section*{Acknowledgments}
We would like to thank the referee for his/her suggestions that helped to improve the manuscript.
The \xmm\ project is supported by the Bundesministerium f\"ur Wirtschaft und
Technologie\,/\,Deutsches Zentrum f\"ur Luft- und Raumfahrt (BMWi/DLR, FKZ 50 OX
0001) and the Max-Planck Society.
We acknowledge the use of publicly available data from the \swift satellite. We thank the \swift team for accepting and carefully scheduling the ToO observations.
MP acknowledges support from the Lyman Spitzer, Jr Fellowship awarded by the Department of Astrophysical Sciences at Princeton University.

\bibliographystyle{mnras}
\bibliography{lmc_long_period}

\end{document}